\documentclass{aa} 
\usepackage{caption}
\usepackage{amsmath,xcolor}
\usepackage{lineno}
\usepackage[utf8]{inputenc}
\usepackage[title]{appendix}%
\usepackage{graphicx}
\usepackage[version=3]{mhchem}
\usepackage{txfonts}
\usepackage{hyperref}
\usepackage{placeins}
\begin{document}

   \title{Investigating the chemical link between H$_2$CO and CH$_3$OH within the CMZ of NGC 253}
    \titlerunning{Multi-transition study of methanol and formaldehyde towards NGC\,253}
    \authorrunning{Huang et al.}


   \author{K.-Y. Huang
          \inst{1}, 
          E. Behrens
          \inst{2,3},
          M. Bouvier
          \inst{1}\fnmsep\thanks{bouvier@strw.leidenuniv.nl},
          S. Viti\inst{1,4,5},
          J. G. Mangum
          \inst{3},
          C. Eibensteiner\inst{3}\thanks{Jansky Fellow of the National Radio Astronomy Observatory}
          }

   \institute{Leiden Observatory, Leiden University, PO Box 9513, 2300 RA Leiden, The Netherlands
        \and
            Department of Astronomy, University of Virginia, P.~O.~Box 400325, 530 McCormick Road, Charlottesville, VA 22904-4325, USA
         \and
              National Radio Astronomy Observatory, 520 Edgemont Road, Charlottesville, VA 22903-2475, USA
              \and 
              Transdisciplinary Research Area (TRA) ‘Matter’/Argelander-Institut f\"ur Astronomie, University of Bonn
              \and
              Physics and Astronomy, University College London, UK
             }

   \date{Submitted: December 2023; accepted: }

 
  \abstract
   {The formaldehyde (H$_2$CO) and methanol (CH$_3$OH) molecules have served as traditional tracers of the star formation process for decades.  Studies of the environments which produce formaldehyde and methanol emission, though, have pointed to significant differences in the physical environments within which each molecule resides. }
   {In this article we aim to investigate the detailed physical and chemical conditions which give rise to formaldehyde and methanol emission in the nearby starburst galaxy NGC\,253. }
   {We employ high spatial ($1.^{\prime\prime}$6 or $\sim28$\,pc) and spectral ($\sim10$\,km/s) imaging of the NGC\,253 central molecular zone (CMZ) from the ALCHEMI Large Program to constrain  radiative transfer models of the dense gas 
   volume density, kinetic temperature, molecular species column density, and source filling factor within eight giant molecular clouds (GMCs). 
   We also measure the relative abundances of the two nuclear spin isomers of CH$_3$OH 
   to investigate its formation history. }
   { The physical and chemical conditions derived 
   clearly indicate that these two molecular species originate from distinct physical environments.  H$_2$CO traces low volume density and high kinetic temperatures, while CH$_3$OH traces high volume density and low kinetic temperatures.  The H$_2$CO abundances are constant, though poorly constrained, within the eight NGC\,253 GMCs analyzed, while the CH$_3$OH abundance shows a radial gradient from low to high values within the NGC\,253 CMZ. }
   {Our findings highlight the complex chemical and physical differentiation of CH$_3$OH and H$_2$CO in the starburst environment of NGC 253. 
   Methanol formation appears to be influenced by warm, dynamic processes rather than cold cloud chemistry, while formaldehyde primarily forms via gas-phase reactions. These results challenge the assumption of a direct chemical link between CH$_3$OH and H$_2$CO and underscores the impact of starburst-driven shocks, turbulence, and cosmic rays on molecular gas chemistry. 
}


   \maketitle
%
\section{Introduction}
The evolution of galaxies is shaped by several key physical and chemical processes within the interstellar medium (ISM). 
Such processes are often tied to star formation and active galactic nuclei (AGN) activities. 
Strong stellar feedback associated with high star-forming rates (SFRs) can trigger outflows of ionized, neutral, and molecular gas. 
The significant energy input from starburst activities makes starburst galaxies prime targets for investigating feedback processes in the ISM \citep{Melioli2004A&A, Krieger2020ApJ}

NGC\,253 is a barred spiral galaxy with an inclination of $76^{\circ}$ \citep{McCormick+2013} that makes it almost edge-on. 
The proximity of NGC\,253 \citep[D $\sim3.5 \pm 0.2$ Mpc,][]{Rekola+2005} also makes it one of the most well studied starburst systems. 
The central molecular zone (CMZ) of NGC\,253 extends approximately $300\times100$ pc \citep{Sakamoto+2011} and encompasses over ten extensively studied giant molecular clouds (GMCs). 
With a SFR of $\sim2 $~M$_{\odot}$yr\textsuperscript{-1} arising from its CMZ \citep[][]{Leroy+2015,Bendo+2015}, which is half of its global star formation activity, NGC\,253 is also classified as a prototype of a nuclear starburst system. 
A large-scale outflow in NGC\,253 has been revealed by multiwavelength observations: in X-rays \citep{Strickland+2000,Strickland+2002}, H$\alpha$ \citep{Westmoquette+2011}, molecular emission \citep{Turner1985,Bolatto+2013,Walter+2017,Krieger+2019}, and dust \citep{Levy+2022}. 
This large-scale outflow is considered a result of the galaxy's starburst activity \citep{McCarthy+1987} due to the lack of AGN signatures \citep{MS+2010,Lehmer+2013}. 
At the more localized scales, shocks and turbulence also contribute mechanical energy to the ISM feedback processes. 

To trace shocks, there are well-known molecular tracers such as silicon monoxide (SiO) and isocyanic acid (HNCO) \citep{sio_MP+1997,Schilke+1997, Huttemeister+1998,Zinchenko+2000,J-S+2008_shocktracers,Martin+2008_HNCO_galactic,hnco+RF+2010}. 
The presence of gas-phase methanol (\ce{CH3OH}) in the ISM of starburst galaxies has also been proposed as an indicator of the shock influence. As the simplest complex organic molecule (COM; \citealt{Herbst_vanDishoeck_2009}), \ce{CH3OH} can form efficiently in cold environments (12-20K) on ice grains with continuous hydrogenation of CO \citep{Fuchs+2009} and H-abstraction reaction of radicals on ice grains \citep{Alvarez+2018, Simons+2020,Santos+2022}. 
The solid-state \ce{CH3OH} can be liberated into gas phase via thermal sublimation or non-thermal desorption processes. 
In the presence of shocks, ice mantle sputtering quickly releases solid-state \ce{CH3OH} into the gas phase. On the other hand, the destruction of \ce{CH3OH} can also take place in the presence of dissociative shock conditions, as proposed by \citet{Suutarinen+2014}. 
Nevertheless, overall, the non-dissociative shocks remain efficient in enhancing gas-phase \ce{CH3OH} in the ISM. 

Formaldehyde (\ce{H2CO}) is an important precursor of \ce{CH3OH} in its formation process on ice grains. 
Unlike methanol, however, \ce{H2CO} can also form in the cold gas phase by reactions of CH$_2$ and CH$_3$ with O$_2$ and O, respectively, depending on the region (Ramal-Olmedo et al. 2021). In warmer regions, H$_2$CO can form by reactions of CH$_3^+$ with OH \citep{Woodall2007AandA}. 
As a slightly asymmetric rotor, \ce{H2CO} has been found to be an excellent tracer of the gas kinetic temperature \citep{Mangum_Wootten_1993}. 

Molecules that host two or more hydrogen atoms such as \ce{H2O}, \ce{H2CO}, and \ce{NH3}, exist in two isomeric forms: ortho- (hydrogen nuclei spin are parallel) and para- (hydrogen nuclei spin are anti-parallel). 
Both \ce{CH3OH} and \ce{H2CO} have two spin isomers. 
For \ce{CH3OH}, they are referred to as E- and A- \ce{CH3OH} owing to the combinations of nuclear spin alignment in the three H-atoms of the methyl (\ce{CH3}) group. 
In A-type \ce{CH3OH}, the nuclear spins of the three protons in the methyl group are parallel. 
For E-type \ce{CH3OH}, one of the protons in the methyl group has an anti-parallel nuclear spin with respect to the others. The E/A ratio can be a probe of the history of methanol formation (see Section 5.3). 
On the other hand, for \ce{H2CO}, the two nuclear spin isomers are referred to as o- and p- \ce{H2CO}. 
Observationally the spin isomers of both molecules have different spectroscopic properties including transition frequencies. 

Both \ce{CH3OH} and \ce{H2CO} have been imaged in NGC\,253. 
In particular, \citet{Mangum+2019} imaged the CMZ of NGC\,253 with multi-transition \ce{H2CO} observations at spatial resolution $1.^{\prime\prime}5-0.^{\prime\prime}4$ and found that, while at large scales ($\sim5.^{\prime\prime}0$) $T_{kin}\gtrsim50$\,K, at smaller scales  ($\lesssim 1.5^{\prime\prime}$) temperatures were at least 300 K, highlighting the importance of high spatial resolution in order to determine the real temperature structure of the central region of this starburst galaxy.  
\ce{CH3OH} was imaged by the ALMA large program, ``ALMA Comprehensive High-resolution Extragalactic Molecular Inventory" \citep[ALCHEMI,][]{ALCHEMI_main_2021}.
Several ALCHEMI studies have already been published and have revealed a high cosmic-ray ionization rate \citep{Holdship+2021_SpectralRadex, Holdship+2022, Harada+2021, Behrens2022ApJ, Behrens2024}, as well as  various types of shocks throughout the CMZ \citep[e.g.][]{Humire+2022, Harada+2022,Huang+2023}. 

The aim of the current study is to investigate both the physical and chemical properties traced by \ce{CH3OH} and \ce{H2CO} in the nearby starburst galaxy, NGC\,253, and investigate whether the two species are chemically linked. 
We present a multi-transition molecular study using \ce{CH3OH} and \ce{H2CO} observations from the ALMA large program, ALCHEMI \citep{ALCHEMI_main_2021}. 
The paper is structured as follows. 
In Section~\ref{sec:obs} we describe the observations and the transition selections for our analysis. 
In Section~\ref{sec:line_emission} we discuss the molecular transition intensity images. 
In Section~\ref{sec:radex} we present our non-LTE radiative transfer analysis in order to constrain the physical conditions of the gas traced by \ce{CH3OH} and \ce{H2CO}. We then explore in Section~\ref{sec:Discussion} the physical origin of the CH$_3$OH and H$_2$CO emission and investigate whether these two molecules are chemically linked.  This discussion includes an assessment of the relative abundances of CH$_3$OH, H$_2$CO, SiO, HNCO, OCS, H$_2$CS, H$_2$S, and CS within eight of the giant molecular clouds (GMCs) of the NGC\,253 CMZ.  We then summarize our findings in Section~\ref{sec:conclusions}. 

\section{Observations and data analysis}
\label{sec:obs}
In this section we summarize the observational setup and the fundamental properties of the ALCHEMI survey data. 
Full details regarding the data acquisition, calibration, and imaging are described in \cite{ALCHEMI_main_2021}. 
ALCHEMI is an ALMA Large Program (project code 2017.1.00161.L and 2018.1.00162.S) 
that imaged the CMZ of NGC\,253 with rest-frequency coverage from 84.2 to 373.2\,GHz, corresponding to the ALMA frequency Bands 3, 4, 5, 6, and 7. 
With wide and thorough spectral scanning across the CMZ of NGC\,253, ALCHEMI provides a comprehensive molecular view of its nucleus. 
A common rectangular area with a size of $50^{\prime\prime} \times 20^{\prime\prime}$ ($850\times340$\,pc) at a position angle of $65^\circ$ was imaged to cover the central nuclear region in NGC\,253. 
A common maximum recoverable angular scale of $15^{\prime\prime}$ was achieved after combining the 12\,m Array and Atacama Compact Array (ACA) measurements at all frequencies. 
The nominal phase center of the observations is $\alpha(\rm ICRS)$ = 00$^h$47$^m$33$^s$.26, $\delta(\rm ICRS)$ = $-25^\circ$17$^\prime$17$^{\prime\prime}.7$. 
The angular and spectral resolution of the image cubes generated from these measurements were $1.^{\prime\prime}6$ \citep[$\sim28$\,pc,][]{ALCHEMI_main_2021} and $\sim10$ km\,s$^{-1}$, respectively. 

From the ALCHEMI data, we extracted the spectral cubes of NGC\,253 for the selected 39 \ce{CH3OH} transitions (Table~\ref{tab:table_obsinfo_ch3oh}) and 10 \ce{H2CO} transitions (Table~\ref{tab:table_obsinfo_h2co}). 
We chose \ce{CH3OH} and \ce{H2CO} transitions based on the following criteria: (1) the transition needs to be bright enough, specifically with signal-to-noise ratio (SNR) greater than 3; and (2) no significant contamination from other molecule(s) involved. In regard to this criterion, we note that the methanol transitions at 165 GHz and 302 GHz are blended with, respectively, an SO$_2$ transition ($\sim$ 27\%) and an NH$_2$OH transition ($\sim$ 23\%), while the H$_2$CO transitions at 140 GHz and 364 GHz suffer from contamination by, respectively, a transition of NH$_2$CN ($\sim$ 25\%) and a transition of SO$_2$ ($\sim$ 25\%). We have tested the impact of these contaminants on our derivation of physical parameters performed in Section 4 and found that the derived physical parameters are unchanged within the uncertainties in each parameter fit.   
Note that we do include transitions that show self blending, including among different spin isomers (see Table~\ref{tab:offsets}).

We note that our compilation of \ce{CH3OH} transitions is distinct from the set of \ce{CH3OH} transitions analyzed by \citet{Humire+2022} which had different transition selection criteria from ours, and focused on masing \ce{CH3OH}. 
Specifically, our list of \ce{CH3OH} transitions does not involve any maser candidates from \citet{Humire+2022}. 

We use \texttt{CubeLineMoment}\footnote{\url{https://github.com/keflavich/cube-line-extractor}} \citep{Mangum+2019} to extract integrated spectral-line intensities from our data cubes. 
\texttt{CubeLineMoment} employs a set of spectral and spatial masks to extract robust integrated intensities for a defined list of target spectral frequencies. 
The spectral masking process of \texttt{CubeLineMoment} uses a chosen bright spectral line in each cube as a velocity tracer of the inspected gas component. 
This makes the definition of velocity span for integration precise while avoiding contamination from adjacent but still separable transitions. 
Final products from the \texttt{CubeLineMoment} analysis include moment 0 (integrated intensity; Jy beam\textsuperscript{-1} $\cdot$ km\,s\textsuperscript{-1}), 1 (average velocity; km\,s\textsuperscript{-1}), and 2 (velocity dispersion; km\,s\textsuperscript{-1}) images masked below a $3\sigma$ threshold (channel-based). 
After making the velocity-integrated maps (moment-0 maps) for all the selected transitions, we converted them from [Jy beam\textsuperscript{-1} $\cdot$ km s\textsuperscript{-1}] to [K $\cdot$ km s\textsuperscript{-1}] for further physical modeling analysis using Equation~\ref{eq:jperk}. 
\begin{table*}[t!]
  \centering
  \caption{\ce{CH3OH} transitions in this study}
  \label{tab:table_obsinfo_ch3oh}
  \begin{tabular}{cc|cccc}
  \hline
    Transition & {} & Rest Frequency 
    & E\textsubscript{u} 
    & $A_{ul}$ & Integrated Intensity (GMC 3)  
    \\
    {} & {} & [GHz]  
    & [K] 
    & [s$^{-1}$] & [K\,km\,s$^{-1}$] 
    \\
    \hline
    E-\ce{CH3OH} & $2_{-1} - 1_{-1}$	   & 96.739360	 
    & 12.5	  
    & 2.558E-06	  
    & 111$\pm$17\\
	E-\ce{CH3OH} & $2_0 - 1_0$	       & 96.744550	 
 & 20.1	  
    & 3.407E-06	  
   & \\
	E-\ce{CH3OH} & $2_1 - 1_1$	       & 96.755500	 
 & 28.0	  
    & 2.624E-06	  
    & \\
	A-\ce{CH3OH} & $2_0 - 1_0$	       & 96.741370	 
 & 7.0	  
    & 3.408E-06	  
    & \\
    \hline
	A-\ce{CH3OH} & $2_{-1} - 1_{-1}$	   & 97.582800	 
 & 21.6	  
    & 2.626E-06	  
    & 8.4$\pm$1.3\\
	\hline
    A-\ce{CH3OH} & $3_1 - 4_0$	       & 107.013830	 
    & 28.3	  
    & 3.066E-06	  
    &6.7$\pm$1.0\\
	\hline
    A-\ce{CH3OH} & $3_1 - 2_1$	       & 143.865790	 
    & 28.3	  
    & 1.069E-05	  
    & 12.7$\pm$1.9 \\
	\hline
    A-\ce{CH3OH} & $3_{-1} - 2_{-1}$	   & 146.368330	 
    & 28.6	  
    & 1.125E-05	  
    & 13.3$\pm$2.0\\
	\hline
	A-\ce{CH3OH} & $2_1 - 3_0$	       & 156.602400	 
 & 21.4	  
    & 8.926E-06	  
    & 13.5$\pm$2.0\\
	\hline
	E-\ce{CH3OH} & $7_0 - 7_{-1}$	   & 156.828520	 
 & 78.1	  
    & 1.878E-05	  
    & 2.8$\pm$0.4\\
	\hline
	E-\ce{CH3OH} & $4_0 - 4_{-1}$	   & 157.246060	 
 & 36.3	  
    & 2.098E-05	  
    &   120$\pm$18   \\
	E-\ce{CH3OH} & $1_0 - 1_{-1}$	   & 157.270830	 
 & 15.4	  
    & 2.206E-05	  
    & \\
	E-\ce{CH3OH} & $3_0 - 3_{-1}$	   & 157.272340	 
 & 27.1	  
    & 2.146E-05	  
    & \\
	E-\ce{CH3OH} & $2_0 - 2_{-1}$	   & 157.276020	 
 & 20.1	  
    & 2.182E-05	  
    & \\
	\hline
	E-\ce{CH3OH} & $1_1 - 1_0$	       & 165.050170	 
 & 23.4	  
    & 2.349E-05	  
    & 76$\pm$11   \\
	E-\ce{CH3OH} & $2_1 - 2_0$	       & 165.061130	 
 & 28.0	  
    & 2.343E-05	  
    & \\
	E-\ce{CH3OH} & $14_6 - 15_5$	   & 165.074350	 
 & 422.4  
    & 5.365E-06	  
    & \\
	E-\ce{CH3OH} & $3_1 - 3_0$	       & 165.099240	 
 & 35.0	  
    & 2.333E-05	  
    & \\
	\hline
	E-\ce{CH3OH} & $4_1 - 4_0$	       & 165.190470	 
 & 44.3	  
    & 2.321E-05	  
    & 22.0$\pm$3.3\\
	\hline
    E-\ce{CH3OH} & $5_1 - 5_0$	       & 165.369340	 
    & 55.9	  
    & 2.308E-05	  
    & 11.5$\pm$1.7\\
	\hline
	E-\ce{CH3OH} & $6_1 - 6_0$	       & 165.678650	 
 & 69.8	  
    & 2.295E-05	  
    & 5.6$\pm$0.9\\
	\hline
	E-\ce{CH3OH} & $3_2 - 2_1$	       & 170.060590	 
 & 36.2	  
    & 2.552E-05	  
    & 30.6$\pm$4.6\\
	\hline
	A-\ce{CH3OH} & $4_{-1} - 3_{-1}$	   & 195.146790	 
 & 38.0	  
    & 2.917E-05	  
    & 14.8$\pm$2.2\\
	\hline
	A-\ce{CH3OH} & $1_1 - 2_0$	       & 205.791270	 
 & 16.8	  
    & 6.277E-05	  
    & 13.5$\pm$2.0\\
	\hline
	A-\ce{CH3OH} & $5_{-1} - 4_{-1}$	   & 243.915790	 
 & 49.7	  
    & 5.966E-05	  
    &12.1$\pm$1.8 \\
	\hline
	E-\ce{CH3OH} & $5_2 - 4_1$	       & 266.838150	 
 & 57.1	  
    & 7.736E-05	  
    & 16.0$\pm$2.4\\
	\hline
	A-\ce{CH3OH} & $6_1 - 5_1$	       & 287.670770	 
 & 62.9	  
    & 1.006E-04	  
    &7.3$\pm$1.1\\
	\hline
    A-\ce{CH3OH} & $6_{-1} - 5_{-1}$   & 292.672890	 
    & 63.7	  
    & 1.060E-04	  
    & 7.6$\pm$1.1\\
	\hline
	E-\ce{CH3OH} & $3_0 - 2_{-1}$	   & 302.369770	 
 & 27.1	  
    & 4.661E-05	  
    & 20.3$\pm$3.0\\
	\hline
	A-\ce{CH3OH} & $1_{-1} - 1_0$	   & 303.366920	 
 & 16.9	  
    & 2.263E-04	  
    &36.0$\pm$5.4\\
	\hline
	A-\ce{CH3OH} & $2_{-1} - 2_0$	   & 304.208350	 
 & 21.6	  
    & 2.116E-04	 
    &44.7$\pm$6.7\\
	\hline
	A-\ce{CH3OH} & $3_{-1} - 3_0$	   & 305.473490	 
 & 28.6	  
    & 1.632E-04	  
    &44.1$\pm$6.6\\
	\hline
	E-\ce{CH3OH} & $3_1 - 2_0$	       & 310.192990	 
 & 35.0	  
    & 8.800E-05	  
    &27.4$\pm$4.1\\
	\hline
	A-\ce{CH3OH} & $6_{-1} - 6_0$	   & 311.852610	 
 & 63.7	  
    & 1.712E-04	  
    &14.0$\pm$2.0\\
	\hline
	A-\ce{CH3OH} & $7_{-1} - 7_0$	   & 314.859530	 
 & 80.1	  
    & 1.749E-04	  
    &8.5$\pm$1.3\\
	\hline
	E-\ce{CH3OH} & $6_2 - 5_1$	       & 315.266860	 
 & 71.0	  
    & 1.183E-04	  
    &9.9$\pm$1.5\\
	\hline
	A-\ce{CH3OH} & $7_1 - 6_1$	       & 335.582020	 
 & 79.0	  
    & 1.628E-04	  
    &3.8$\pm$0.6\\
	\hline
	A-\ce{CH3OH} & $1_1 - 0_0$	       & 350.905100	 
 & 16.8	  
    & 3.315E-04	  
    & 33.2$\pm$5.0\\
	\hline
	E-\ce{CH3OH} & $4_1 - 3_0$	       & 358.605800	 
 & 44.3	  
    & 1.319E-04	  
    & 21.6$\pm$3.2\\
	\hline
  \end{tabular}\\
  \tablefoot{Frequencies and spectroscopic parameters have been extracted from the JPL catalogue \citep{Pickett+1998}. The available data for CH$_3$OH are from \cite{Xu+2008}. Note that only transitions covered within $\pm 150$ km s\textsuperscript{-1} with respect to the main transition are listed as shown with vertical lines in Figure \ref{fig:Spec_I} and \ref{fig:Spec_II} - these are considered to be the "contributors" of the velocity-integrated intensities measured from the moment-0 maps. It is worth noting that around 165 GHz, the methanol transition with $E_{\text{u}}=422$ K will only contribute significantly if $T_{\text{kin}}> 100$ K. The main line refers to transitions with the lowest $E_{u}$ in each frequency group.  Measured integrated intensity values and their uncertainties for GMC 3 are provided as a reference. Integrated intensities measured toward the GMCs listed in Table~\ref{tab:GMC_locations} are available as a machine-readable table \href{https://doi.org/10.5281/zenodo.15442733}{online}.}
\end{table*}
\begin{table*}[t!]
  \centering
  \caption{List of \ce{H2CO} transitions in this study}
  \label{tab:table_obsinfo_h2co}
  \begin{tabular}{cc|cccc}
  \hline
    Transition & {} & Rest Frequency 
    & E\textsubscript{u} 
    & $A_{ul}$ & 
    Integrated Intensity (GMC 3)\\
    {} & {} & [GHz]  
    & [K] 
    & [s$^{-1}$] &
    [K\,km\,s$^{-1}]$\\
    \hline
    o-\ce{H2CO} & $2_{12} - 1_{11}$	   & 140.839500	   
    & 21.9	   
    & 5.304E-05	   & 
    104$\pm$16\\
	\hline
	o-\ce{H2CO} & $3_{12} - 2_{11}$	   & 225.697800	   
 & 33.4	   
    & 2.772E-04	   & 
    46.0$\pm$6.9\\
	\hline
	o-\ce{H2CO} & $4_{14} - 3_{13}$	   & 281.526900	   
 & 45.6	   
    & 5.883E-04	   & 
    42.5$\pm$6.4\\
	\hline
	p-\ce{H2CO} & $4_{23} - 3_{22}$	   & 291.237800	   
 & 82.1	   
    & 5.210E-04	   & 
    7.1$\pm$1.1\\
	\hline
	o-\ce{H2CO} & $4_{32} - 3_{31}$	   & 291.380500	   
 & 140.9	   
    & 3.044E-04	   & 
    15.6$\pm$2.3\\
	o-\ce{H2CO} & $4_{31} - 3_{30}$	   & 291.384300	   
 & 140.9	   
    & 3.044E-04	   & 
    \\
	\hline
	p-\ce{H2CO} & $4_{22} - 3_{21}$	   & 291.948100	   
 & 82.1	   
    & 5.249E-04	   & 
    8.8$\pm$1.4\\
	\hline
	o-\ce{H2CO} & $5_{15} - 4_{14}$	   & 351.768600	   
 & 62.5	   
    & 1.202E-03	   & 
    27.6$\pm$4.1\\
	\hline
	p-\ce{H2CO} & $5_{24} - 4_{23}$	   & 363.945900	   
 & 99.5	   
    & 1.165E-03	   & 
    5.8$\pm$0.9\\
	\hline
	p-\ce{H2CO} & $5_{23} - 4_{22}$	   & 365.363400	   
 & 99.7	   
    & 1.179E-03	   & 
    4.7$\pm$0.7\\
	\hline
  \end{tabular}\\
  \tablefoot{Frequencies and spectroscopic parameters have been extracted from CDMS \citep{Muller+2005, Endres+2016}. The available data for H$_2$CO are from \cite{Muller+2017}. Note that only transitions covered within $\pm 150$ km s\textsuperscript{-1} with respect to the main transition are listed as shown with vertical lines in Figure \ref{fig:Spec_II} - these are considered to be the "contributors" of the velocity-integrated intensities measured from the moment-0 maps. The main line refers to transitions with the lowest $E_{u}$ in each frequency group.  Measured integrated intensity values and their uncertainties for GMC 3 are provided as a reference. Integrated intensities measured toward the GMCs listed in Table~\ref{tab:GMC_locations} are available as a machine-readable table \href{https://doi.org/10.5281/zenodo.15442733}{online}.}
\end{table*}
\section{Molecular line emission}
\label{sec:line_emission}
\begin{figure}
    \centering
    \includegraphics[width=\linewidth]{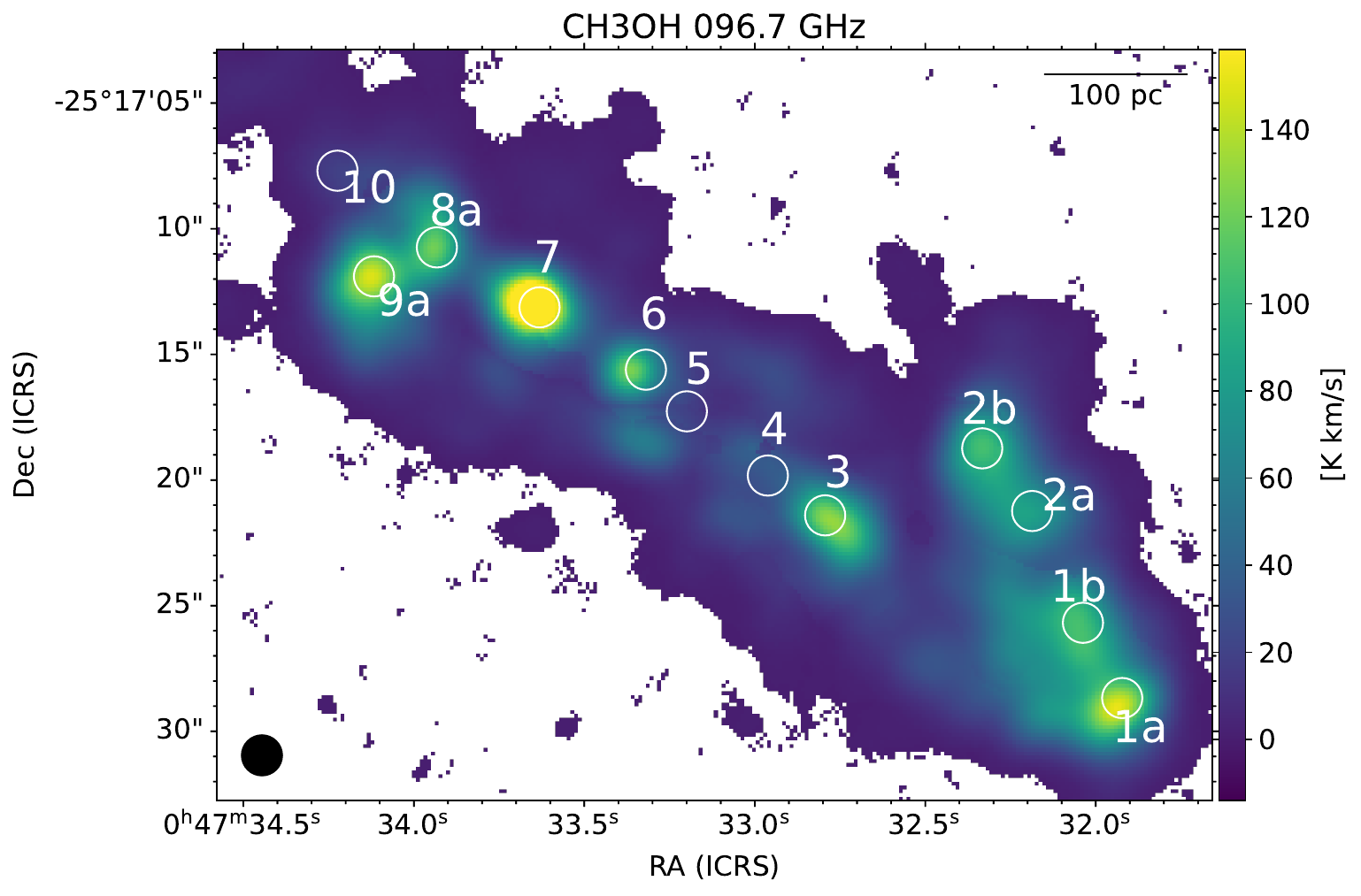} \\
    \includegraphics[width=\linewidth]{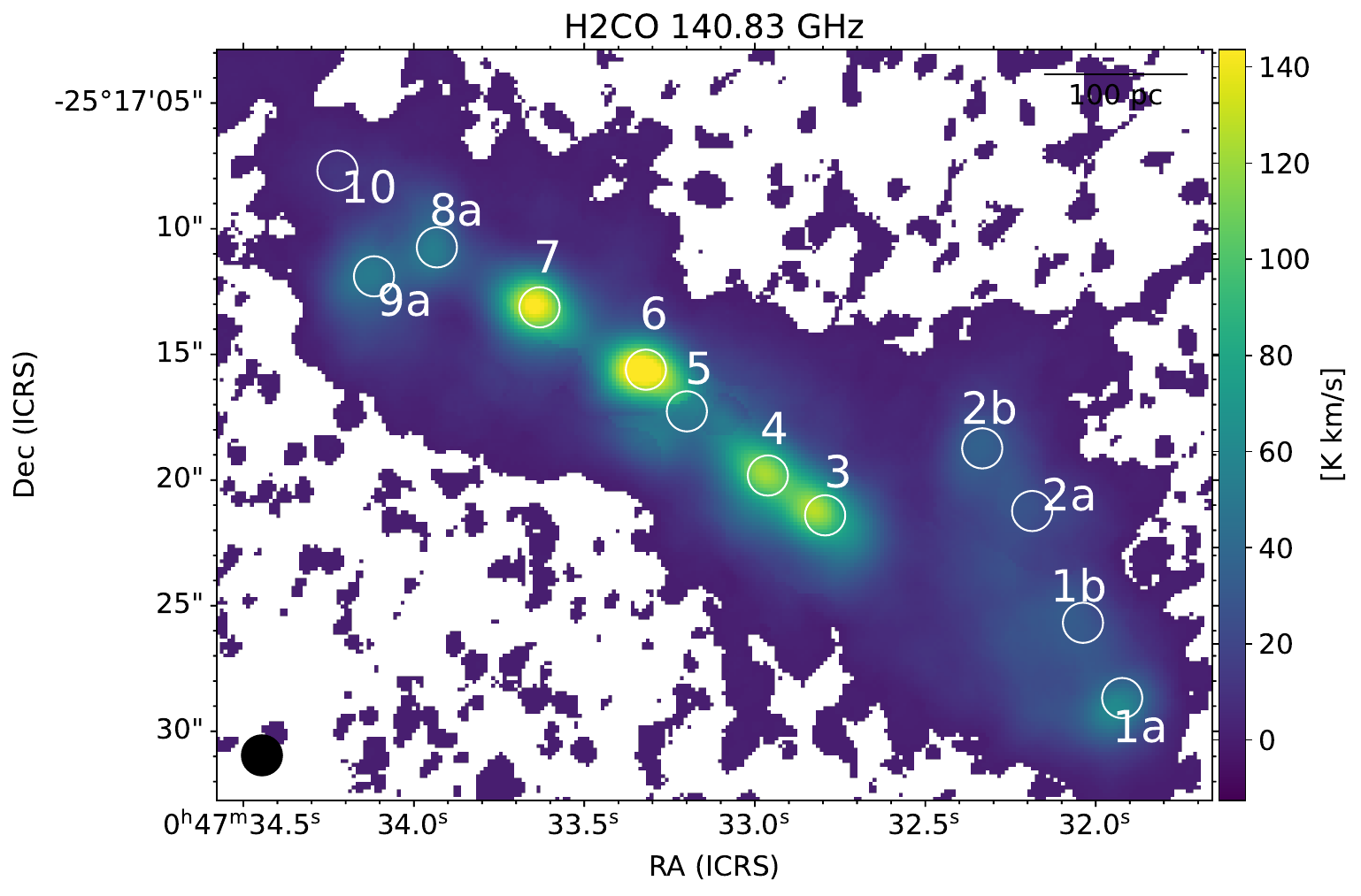}
    \caption{Moment-0 maps of all the \ce{CH3OH} transitions in the 96 GHz group and the \ce{H2CO} $2_{12}-1_{11}$ transition near 140 GHz showing integrated emission with SNR $>$ 3. White circles indicate locations of GMCs studied in \cite{Huang+2023}. Black circle in bottom left corner represents the 1.$^{\prime\prime}6$ ($\sim28$ pc) ALCHEMI beam.}
    \label{fig:Maps_m3_mom0-I}
\end{figure}

In Fig. \ref{fig:Maps_m3_mom0-I} we present the velocity-integrated transition intensity images from the \ce{CH3OH} 96.7 GHz transition group and \ce{H2CO} $2_{12} - 1_{11}$ transition as representative cases for each species (see Appendix~\ref{sec:mom0_maps} for the integrated intensity maps for all transitions listed in Tables~\ref{tab:table_obsinfo_ch3oh} and \ref{tab:table_obsinfo_h2co}). 
We also show the representative spectra extracted from GMC 6 for all transitions analyzed in this work in Appendix~\ref{sec:spectra} (Fig. \ref{fig:Spec_I} and \ref{fig:Spec_II}).  We convert the integrated emission from Jy\,beam$^{-1}$\,km\,s$^{-1}$ to K\,km\,s$^{-1}$ using the following expression
\begin{equation}
    I(\text{K}) = 13.59\left(\dfrac{300\text{GHz}}{\nu}\right)^2 \times \left(\dfrac{1^{\prime\prime}}{\theta_{max}}\right)\left(\dfrac{1^{\prime\prime}}{\theta_{min}}\right)I(\text{Jy}),
    \label{eq:jperk}
\end{equation}
where $\nu$ is the rest frequency of the line, $\theta_{max}$ and $\theta_{min}$ are the major and minor axes, respectively, of our Gaussian beam, and $I$ is the integrated intensity.

The molecular line emission presented in Fig. \ref{fig:Maps_m3_mom0-I} shows several bright regions across the CMZ of NGC\,253. 
Overall they match well with the line emission spatial distribution reported by prior ALCHEMI studies with other molecular species - including \ce{C2H} \citep{Holdship+2022}, \ce{HCO+} and \ce{HOC+} \citep{Harada+2021}, \ce{H3O+} and \ce{SO} \citep{Holdship+2022}, \ce{HOCO+} \citep{Harada+2022}, \ce{HCN} and \ce{HNC} \citep{Behrens2022ApJ,Behrens2024}, \ce{HNCO} and \ce{SiO} \citep{Huang+2023}, sulfur-bearing species \citep{Bouvier+2024}, and \ce{CO} and their isotopologues \citep{Butterworth+2024}. 

We note that absorption features in GMC 5 - due to the strong continuum in the background toward the center of the galaxy and potential self-absorption in the optically thick regime - have been reported from multiple studies with different species \citep[e.g., ][]{Meier+2015_hncosio_253,Humire+2022}. 
Such features are also seen in our data. To interpret the intensity measurements from this region requires an advanced radiative-transfer model which includes strong background continuum emission.  Such a modeling effort is beyond the scope of the present study.  
Additionally, in GMC 10, there is a lack of detection in most transitions. 
As a result, we do not discuss these two regions (GMCs 5 and 10) any further.
The positions of all the remaining GMCs are listed in Table \ref{tab:GMC_locations}. 
These 8 GMC regions are also marked as white circles on both representative CH$_3$OH and H$_2$CO moment-0 images shown in Fig. \ref{fig:Maps_m3_mom0-I}. 

\begin{table}[ht!]
  \centering
  \caption{The 8 selected NGC\,253 GMC positions described in Sect. \ref{sec:line_emission}. }
  \label{tab:GMC_locations}
  \begin{tabular}{ccc}
  \hline
    {GMC} & {R.A.(ICRS)} & {Dec.(ICRS)} \\
    {}& {(00$^h$ 47$^m$)} & {($-25^\circ$ 17$^\prime$)}\\
    \hline
    \hline
    GMC\,1a & 31$^s$.9344 & 28$^{\prime\prime}$.822 \\
    GMC\,2b & 32$^s$.3449 & 18$^{\prime\prime}$.886 \\
    GMC\,3 & 32$^s$.8056 & 21$^{\prime\prime}$.552 \\
    GMC\,4 & 32$^s$.9736 & 19$^{\prime\prime}$.968 \\
    GMC\,6 & 33$^s$.3312 & 15$^{\prime\prime}$.756 \\
    GMC\,7 & 33$^s$.6432 & 13$^{\prime\prime}$.272 \\
    GMC\,8a & 33$^s$.9443 & 10$^{\prime\prime}$.888 \\
    GMC\,9a & 34$^s$.1287 & 12$^{\prime\prime}$.040 \\
    \hline
  \end{tabular}
\end{table}

\section{Non-LTE radiative transfer modeling analysis}
\label{sec:radex}
We performed non-LTE radiative transfer modeling to characterize the average gas properties traced by \ce{CH3OH} and \ce{H2CO}  over beam-sized 
($1.^{\prime\prime}6$ or 28\,pc) structures. 
Given the plethora of transitions we have measured for both species, spanning a wide range in excitation energy,  radiative transfer analysis can be used to properly constrain the gas physical characteristics. 
We use the radiative transfer code \texttt{RADEX} \citep{radex_vandertak_2007} via the Python package \texttt{SpectralRadex}\footnote{https://spectralradex.readthedocs.io} \citep{Holdship+2021_SpectralRadex}. 
The collisional excitation rates used for \ce{CH3OH} are from \citet{Dagdigian_2023} while the \ce{H2CO} rates are from \citet{H2CO_mole_data}.  These rates were obtained from the Excitation of Molecules and Atoms for Astrophysics (EMAA)\footnote{\url{https://emaa.osug.fr/}} and the LAMDA databases \citep{Schoier2005A&A}, respectively. For kinetic temperature priors below or above the lower or upper limits to the kinetic temperatures at which collisional excitation rates are provided (3 to 250 and 10 to 300\,K for CH$_3$OH and H$_2$CO, respectively), \texttt{SpectralRadex} uses the minimum or maximum kinetic temperature at which collisional excitation rates are provided.

One should note that there is a noticeable difference in the derived E-/A- spin isomer column density ratio (EAR) for CH$_3$OH when using the old (from \citealt{CH3OH_mol_data_2010}) and the new (from \citealt{Dagdigian_2023}) collisional excitation rates. With the \cite{CH3OH_mol_data_2010} rates, the average range for the best EAR was 0.85--2.84 whilst with the \cite{Dagdigian_2023} rates, the range is 0.25--1.26. By comparing the brightness temperatures for the $3_0-2_0$A and $2_{-1}-1_{-1}$E lines of \ce{CH3OH} at 145.103 GHz and 96.739 GHz, respectively, \cite{Dagdigian_2023} showed that the largest difference in both quantities occurred for the E-type transition at low temperature (10 K; see their Fig. 5), with the brightness temperature calculated with the new rates being lower (by maximum 10\%) than with the previous rates, in particular for $n_{\mathrm{H2}}\geq 10^4$ cm$^{-3}$. The difference is less important at 100 K. Although the comparison has been made for one E-type transition, this difference could explain the change in the derived EAR between the old and new excitation rates that we found. The choice of the rates used for \ce{CH3OH} thus seem more critical if one aims  to derive the EAR.
 
We coupled radiative transfer modeling with the \texttt{Nautilus} sampler \citep{nautilus}\footnote{As a side note, we find that \texttt{Nautilus} is substantially more efficient than \texttt{UltraNest} \citep{ultranest21} \citep[used in][]{Holdship+2022,Behrens2022ApJ,Huang+2023}, while providing indistinguishable results. } for Bayesian posterior and evidence estimation. 
We then perform the Bayesian inference of the parameter probability distributions. 
We assume priors of uniform or log-uniform distribution within the determined ranges (listed in Table \ref{tab:table_prior}). 
We also assume that the uncertainty on our measured intensities is Gaussian so that our likelihood is given by $P(\theta | d) \sim \exp(-\frac{1}{2}\chi^2),$ where $\chi^2$ is the chi-squared statistic between our measured intensities and the \texttt{RADEX} predictions based on a set of parameters $\theta$. 
This allows us to assess the influence of the physical conditions in the excitation of the transitions analyzed. Note that for each blending group, we model the integrated intensities of the transitions that make up the blending group separately, sum up the emission from each of these lines, and then compare this summed value to the total measured blended intensity for that group.

The following physical parameters are explored: the gas volume density ($n_{H2})$, gas kinetic temperature ($T_{kin}$), the species column density of both spin isomers of each molecule (N(E-CH$_3$OH) and N(A-CH$_3$OH) for \ce{CH3OH} and N(o-H$_2$CO) and N(p-H$_2$CO) for \ce{H2CO}), and the beam filling factor ($\eta_{ff}=\frac{\theta^{2}_{S}}{\theta^{2}_{MB}+\theta^{2}_{S}}$).
The prior ranges and distributions are given in Table \ref{tab:table_prior}. 

We list the best fit for all the explored parameters in Tables \ref{tab:Bayesian_gasphys_ch3oh} and \ref{tab:Bayesian_gasphys_h2co}, and 
present the inference results from a representative GMC (GMC 3) for both \ce{CH3OH} and \ce{H2CO} in Fig. \ref{fig:Baye_GMC3}. The rest of the inference results can be found in Appendix~\ref{sec:corners}. In the following subsections we discuss our findings. 
\begin{figure*}
  \centering
  \includegraphics[scale=0.5]{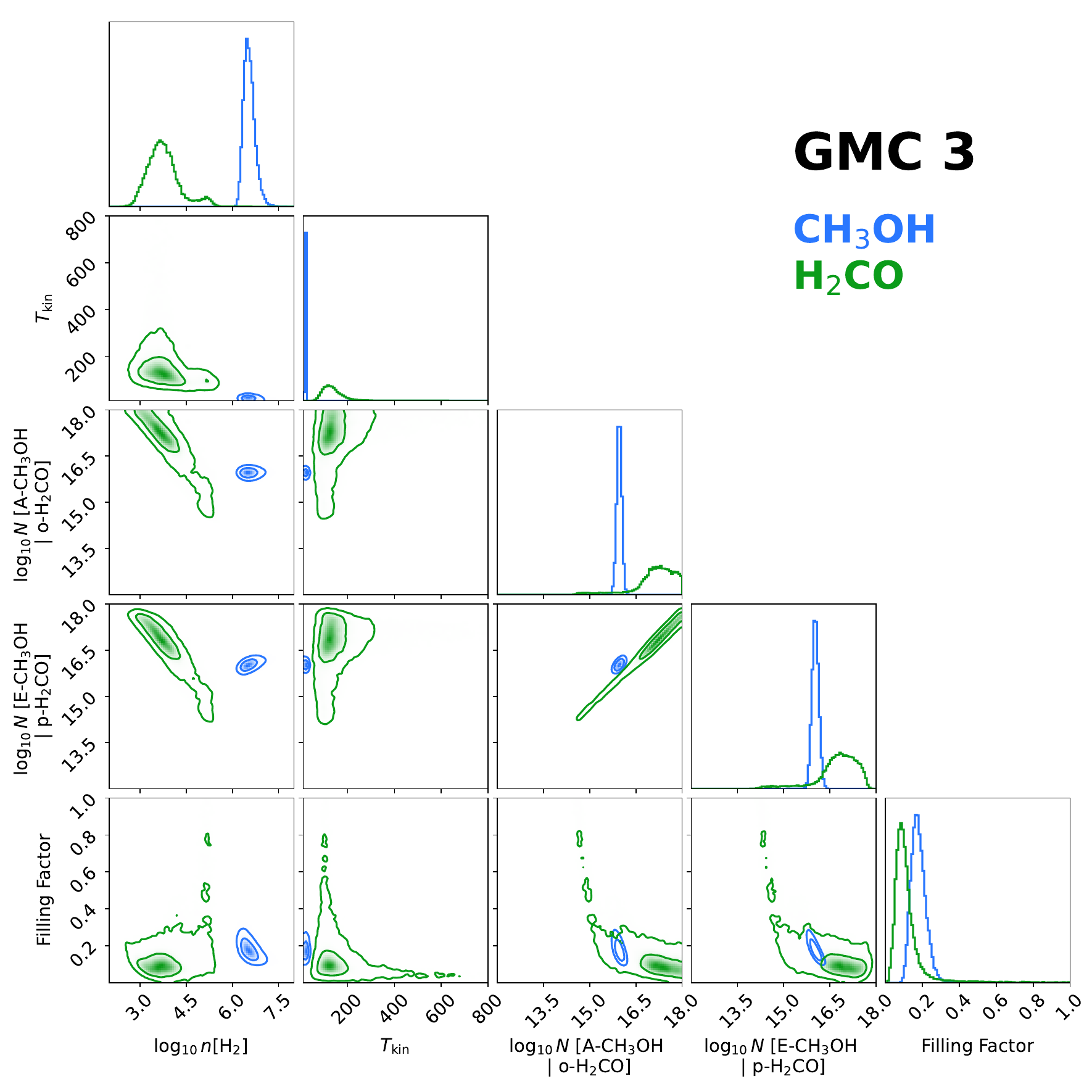}
  \caption{The inference results from GMC 3 with both \ce{CH3OH} (blue) and \ce{H2CO} (green).}
  \label{fig:Baye_GMC3}
\end{figure*}

\begin{table}[ht!]
  \centering
  \caption{Prior ranges adopted for our parameter space explored in the \texttt{RADEX}-Bayesian inference process described in Sect. \ref{sec:radex}. }
  \label{tab:table_prior}
  \begin{tabular}{c|cc}
  \hline
    Variable  & Range & Distribution type\\
    \hline
    Gas density, $n_{\rm H2}$ [cm\textsuperscript{-3}] & $10^{2}-10^{8}$ & Log-uniform\\
    Gas temperature, $T_{\rm kin}$ [K] & $10-800$ & Uniform\\
    $N$(A-\ce{CH3OH}) [cm\textsuperscript{-2}] & $10^{12}-10^{18}$ & Log-uniform\\
    $N$(E-\ce{CH3OH}) [cm\textsuperscript{-2}] & $10^{12}-10^{18}$ & Log-uniform\\
    $N$(o-\ce{H2CO}) [cm\textsuperscript{-2}] & $10^{12}-10^{18}$ & Log-uniform\\
    $N$(p-\ce{H2CO}) [cm\textsuperscript{-2}] & $10^{12}-10^{18}$ & Log-uniform\\
    Beam filling factor, $\eta_{ff}$ & $0.0-1.0$ & Uniform\\
    \hline
  \end{tabular}
\end{table}
\renewcommand{\arraystretch}{1.5}

\begin{table*}
    \centering
    \caption{Inferred gas properties traced by \ce{CH3OH} from the Bayesian inference processes over the GMC regions investigated.  }
    \label{tab:Bayesian_gasphys_ch3oh}
    \resizebox{\linewidth}{!}{\begin{tabular}{c|ccccc|ccc}
        \hline
		GMC & log(n\textsubscript{H2}) & $T_{\text{kin}}$ & log(N\textsubscript{A-CH3OH}) & log(N\textsubscript{E-CH3OH}) & $\eta_{ff}$ & EAR\textsubscript{best} & EAR range & X\textsubscript{CH3OH} range \\ 
		\hline
	   1a & $4.22^{+0.11}_{-0.11}$ & $76.06^{+11.44}_{-9.69}$ & $16.60^{+0.06}_{-0.06}$ & $16.32^{+0.13}_{-0.11}$ & $0.08^{+0.01}_{-0.01}$ & {0.52} & {[0.35, 0.81] } & $1.2\times10^{-8}-1.4\times10^{-7}$\\
        2b & $5.18^{+0.12}_{-0.09}$ & $11.14^{+0.57}_{-0.55}$ & $16.40^{+0.10}_{-0.14}$ & $16.38^{+0.16}_{-0.19}$ & $0.09^{+0.01}_{-0.01}$ & {0.95} & {[0.49, 1.91]}& $1.2\times10^{-8}-1.2\times10^{-7}$\\
	   3  & $6.53^{+0.19}_{-0.16}$ & $19.42^{+0.98}_{-0.93}$ & $15.96^{+0.08}_{-0.08}$ & $16.02^{+0.11}_{-0.11}$ & $0.18^{+0.04}_{-0.03}$ & {1.15} & {[0.74, 1.78]} & $1.8\times10^{-9}-6.1\times10^{-9}$\\
        4  & $6.89^{+0.21}_{-0.17}$ & $26.79^{+1.55}_{-1.37}$ & $14.99^{+0.18}_{-0.11}$ & $15.06^{+0.18}_{-0.10}$ & $0.73^{+0.19}_{-0.25}$ & {1.17} & {[0.62, 2.29] } & $4.0\times10^{-10}-2.2\times10^{-9}$\\
        6  & $6.86^{+0.25}_{-0.19}$ & $21.49^{+1.10}_{-1.01}$ & $15.76^{+0.12}_{-0.15}$ & $15.86^{+0.14}_{-0.16}$ & $0.28^{+0.11}_{-0.07}$ & {1.26} & {[0.66, 2.45]} & $9.3\times10^{-10}-5.3\times10^{-9}$\\
        7  & $5.33^{+0.05}_{-0.05}$ & $23.39^{+0.99}_{-0.96}$ & $16.64^{+0.05}_{-0.05}$ & $16.25^{+0.07}_{-0.07}$ & $0.12^{+0.01}_{-0.01}$ & {0.41} & {[0.31, 0.54]} & $2.8\times10^{-8}-1.4\times10^{-7}$\\
        8a & $4.95^{+0.06}_{-0.06}$ & $29.47^{+2.05}_{-1.74}$ & $16.36^{+0.06}_{-0.05}$ & $15.75^{+0.08}_{-0.08}$ & $0.09^{+0.01}_{-0.01}$ & {0.25} & {[0.18, 0.33]} & $3.7\times10^{-9}-8.8\times10^{-8}$\\
        9a & $4.82^{+0.07}_{-0.07}$ & $24.87^{+1.33}_{-1.22}$ & $16.33^{+0.07}_{-0.07}$ & $15.89^{+0.10}_{-0.10}$ & $0.14^{+0.01}_{-0.01}$ & {0.36} & {[0.25, 0.54]} & $8.0\times10^{-9}-1.3\times10^{-7}$\\
		\hline
        Range & $4.11-7.11$ & $10.59-87.5$ & $14.88-16.66$ & $14.96-16.54$ & $0.07-0.92$ & $0.25-1.26$ & [$0.18-2.45$] &  $4.0\times10^{-10} - 1.4\times10^{-7} $\\
        \hline
    \end{tabular}}\\
    \tablefoot{The "EAR" stands for E-/A- spin isomer column density ratio of \ce{CH3OH}. In the column "EAR\textsubscript{best}" we list the best-estimate EAR in each GMC region based on the inferred column density of A-\ce{CH3OH} and E-\ce{CH3OH} respectively. In the column "EAR range" we list the range of possible EAR subtended by the uncertainties of the inferred E- and A- \ce{CH3OH} column densities.  }
\end{table*}

\begin{table*}
    \centering
    \caption{Inferred gas properties traced by \ce{H2CO} from the Bayesian inference processes over the GMC regions investigated. }
    \label{tab:Bayesian_gasphys_h2co}
    \begin{tabular}{c|ccccc|c}
        \hline
		GMC & log(n\textsubscript{H2}) & $T_{\text{kin}}$ & log(N\textsubscript{o-H2CO}) & log(N\textsubscript{p-H2CO}) & $\eta_{ff}$ &  X\textsubscript{H2CO} range\\ 
		\hline
		1a & $4.05^{+0.77}_{-0.65}$ & $95.94^{+27.58}_{-17.87}$ & $15.99^{+0.94}_{-0.99}$ & $15.13^{+0.91}_{-0.97}$ & $0.21^{+0.27}_{-0.11}$ &  $1.4\times10^{-10}-1.4\times10^{-6}$\\
        2b & {- -} & $83.91^{+18.34}_{-13.78}$ & {- -} & {- -} & {- -} & {- -}\\
		3  & $3.71^{+0.49}_{-0.43}$ & $129.71^{+49.30}_{-32.99}$ & $17.24^{+0.49}_{-0.56}$ & $16.85^{+0.50}_{-0.58}$ & $0.10^{+0.06}_{-0.03}$ &  $2.1\times10^{-9}-1.7\times10^{-7}$\\
        4  & $3.89^{+0.39}_{-0.37}$ & $266.73^{+316.23}_{-122.64}$ & $17.43^{+0.37}_{-0.50}$ & $16.97^{+0.40}_{-0.48}$ & $0.07^{+0.04}_{-0.03}$  & $1.4\times10^{-9}-9.4\times10^{-8}$\\
        6  & $3.86^{+0.37}_{-0.36}$ & $> 200$                      & $17.56^{+0.30}_{-0.46}$ & $17.09^{+0.35}_{-0.47}$ & $0.07^{+0.05}_{-0.03}$ & $1.8\times10^{-9}-1.2\times10^{-7}$\\
        7  & $3.94^{+0.43}_{-0.46}$ & $107.36^{+28.68}_{-21.35}$ & $16.20^{+0.68}_{-0.51}$ & $15.64^{+0.66}_{-0.50}$ & $0.41^{+0.26}_{-0.16}$ & $5.7\times10^{-9}-1.1\times10^{-6}$\\
        8a & $3.88^{+0.65}_{-0.65}$ & $94.12^{+22.82}_{-16.30}$ & $15.69^{+1.02}_{-0.69}$ & $15.00^{+0.99}_{-0.71}$ & $0.36^{+0.33}_{-0.19}$ &  $3.3\times10^{-10}-1.2\times10^{-6}$\\
        9a & $4.14^{+0.61}_{-0.55}$ & $79.91^{+17.51}_{-12.68}$ & $15.83^{+0.68}_{-0.79}$ & $14.97^{+0.69}_{-0.81}$ & $0.22^{+0.25}_{-0.10}$ &  $1.7\times10^{-10}-3.1\times10^{-7}$\\
		\hline
        Range & $3.50-4.75$ & $\geq67.23$ &$15.00-17.86$ & $14.16-17.44$ & $0.04-0.67$ & $1.4\times10^{-10}-1.4\times10^{-6}$  \\
        \hline
    \end{tabular}\\
\end{table*}

\renewcommand{\arraystretch}{1.0}
\subsection{Properties of the gas as traced by CH$_3$OH and H$_2$CO}
\label{sec:radex_phys}
In general \ce{CH3OH} traces dense gas components, with gas volume density ranging from $\sim 10^{5} - 10^{7}$ cm\textsuperscript{-3}. 
The methanol in the inner GMCs (GMCs 3, 4, 6, 7) traces the densest gas components (n\textsubscript{H2} $\gtrsim 4\times 10^{5} - 10^7$ cm\textsuperscript{-3}) compared to the rest of the GMCs in the outer regions (n\textsubscript{H2} $\leq 2\times10^{5}$ cm\textsuperscript{-3}). 
The gas temperatures probed by \ce{CH3OH} seem to be surprisingly low, with the lowest temperature being $\sim11$\,K in GMC 2b 
and not exceeding 30\,K in all GMCs except 1a. In this region, we infer a gas temperature of $\sim76$\,K, more than twice the temperature inferred from CH$_3$OH in all other regions across NGC\,253. We note that using the more recent \ce{CH3OH} collisional excitation rates from \cite{Dagdigian_2023} resulted in a higher kinetic temperature, whereas the rates calculated by \cite{CH3OH_mol_data_2010} yielded a temperature of $\sim36$\,K. However, both \cite{Huang+2023} and \cite{Tanaka2024} also identified high ($>100$\,K) gas temperatures in GMC 1a using HNCO and a selection of dense gas tracers, respectively. \cite{Gorski2017} also noted the presence of an expanding shell of gas near GMC 1a, and \cite{Tanaka2024} suggest this could indicate that the gas near GMC 1a is being shocked and mechanically heated by a supernova remnant.

While \ce{CH3OH} seems to largely trace cold gas, \citet{Rico-Villas+2020} used vibrationally excited transitions of HC$_3$N to derive gas kinetic temperatures of $\gtrsim 100$\,K across all the super star clusters (SSCs) in the inner GMCs of NGC\,253, with gas density n\textsubscript{H2} $\gtrsim 10^{6}$ cm\textsuperscript{-3}. It is possible that $v=0$ CH$_3$OH is emitted from a lower temperature gas, especially if we are seeing a post-shocked gas whereby methanol has been released (e.g. via shocks) in the gas phase from ices; nevertheless it is worth also exploring the possibility that methanol is subthermally excited.
We therefore inspected the excitation conditions given the low gas temperatures probed by \ce{CH3OH}. 
Specifically, we compared the collisional and radiative excitation rate \citep{Goldsmith+2012} based on the on-the-spot measurements of opacity, gas density, and gas temperature in each GMC.
We find that overall only about half of the \ce{CH3OH} transitions may be undergoing sub-thermal excitation conditions. 
Moreover, at the low temperature probed by \ce{CH3OH}, the molecule is either expected to still be frozen on the dust (if methanol formed on the ices) or to freeze fairly quickly (if methanol was formed in the gas phase when the gas was  warmer\footnote{At n\textsubscript{H2} $\sim 10^{7}$ cm\textsuperscript{-3} the freeze-out time scale $\sim 10^{9}/\rm{n_{H2}} \sim 10^2$ yrs; this can be longer if one considers non-thermal desorption \citep{Roberts2007MNRAS}}).  Alternatively, we can invoke either the presence of periodic shocks and/or desorption via cosmic rays, as a non-thermal desorption process that releases the \ce{CH3OH} from the ices. 

In contrast, \ce{H2CO} probes gas components of lower density compared to \ce{CH3OH}. In fact, the gas density traced by \ce{H2CO} is surprisingly homogeneous across all GMCs (n\textsubscript{H2} $\sim 10^4$ cm\textsuperscript{-3}). 
On the other hand, \ce{H2CO} traces warmer gas, with  $T_{\text{kin}}\gtrsim 100$\,K across all GMCs, with the highest $T_{\text{kin}}$ in GMCs 4 and 6. 
We also compared the collisional and radiative excitation rates based on on-the-spot measurements of opacity, gas density, and gas temperature in each GMC for \ce{H2CO}. 
Surprisingly the majority of \ce{H2CO} transitions possess sub-thermal excitation conditions. 
The fact that we do not see H$_2$CO in the cold component traced by methanol implies that, if methanol is indeed sublimated by non-thermal desorption processes,  solid H$_2$CO is not abundant,  making it unlikely that their formation paths are chemically connected. 
We therefore suggest that the H$_2$CO we observe is mainly formed in the gas phase, via reactions of CH$_3$ and atomic oxygen \citep{Loomis2015ApJL, Guzman2013A&A}.  If these two species were to both form on the ices, then once non-thermally desorbed we would expect them to be co-spatial.

\subsection{Comparison with existing measurements of CH$_3$OH and H$_2$CO towards NGC\,253}
\citet{Humire+2022} estimated molecular column densities and excitation temperatures ($T_{ex}$) of E-/A- \ce{CH3OH} via an LTE analysis assuming optically thin conditions. Comparing our CH$_3$OH column density measurements (Table~\ref{tab:Bayesian_gasphys_ch3oh}) to those reported by \cite{Humire+2022}, with the exception of GMC 4, our derived column densities of E- and A- \ce{CH3OH} are at least 1-2 orders of magnitude higher. This discrepancy can be explained by the fact that their LTE analysis assumed a filling factor of 1. As can be seen in Table 5, all of our best fits derive a much smaller filling factor than 1, with the exception of GMC 4 where the best fit filling factor is $\sim$0.7. 
We note that the $T_{ex}$ from \citet{Humire+2022} is consistent with our kinetic temperature using non-LTE radiative transfer modeling. 
Similarly, we also compare our inference of \ce{H2CO} column density as well as our inferred kinetic temperatures with those derived by \citet{Mangum+2019} where optically thin conditions were also assumed\footnote{ \citet{Mangum+2019} assumed an excitation temperature T = 150\,K and a beam filling factor of 1}. 
Our derived column densities of o- and p- \ce{H2CO} are at least 1-2 orders of magnitude higher than the prediction by \citet{Mangum+2019}, again probably due to the fact that they assumed a beam filling factor of 1. 
Our derived kinetic temperatures are all greater than 80 K, consistent with the high kinetic temperature derived by \citet{Mangum+2019}. 


\subsection{Derivation of H$_2$CO and CH$_3$OH abundances}\label{sec:abundances}
To derive the methanol and formaldehyde abundances, we used the H$_2$ column densities from \cite{Mangum+2019}. In particular, we used the values of $N_{H2}$ derived from the continuum emission at 220 GHz (see their Table 5). Since the beam of their observations is slightly smaller compared to ours ($1.56'' \times 0.89 '' $ versus $1.6'' \times 1.6''$), we rescaled their H$_2$ column densities to match our angular resolution. This resulted in multiplying their $N_{H2}$ by a factor of 0.63. Then, as the column densities derived for H$_2$CO and CH$_3$OH correspond to a certain beam-filling factor, we applied a beam filling factor to the scaled H$_2$ column densities for each region, using the values of $\eta_{ff}$ from Table~\ref{tab:Bayesian_gasphys_ch3oh}.  In \cite{Mangum+2019} the H$_2$ column densities being only available for the regions GMC\,3 to GMC\,7, we used the $N_{H2}$ values from \cite{Harada+2022} for the other regions, i.e., for GMC\,1a, GMC\,2b, GMC\,8a, and GMC\,9a. For these regions, we used $N_{H2}=(3-10)\times10^{22}$ cm$^{-2}$ (see their Figure 4c). As the observations in \cite{Harada+2022} are also ALCHEMI observations, we did not rescale $N_{H2}$. However, we did apply the beam filling factors from Table~\ref{tab:Bayesian_gasphys_h2co} to the H$_2$ column densities for each region. 

\begin{figure*}
    \centering
\includegraphics[width=\linewidth]{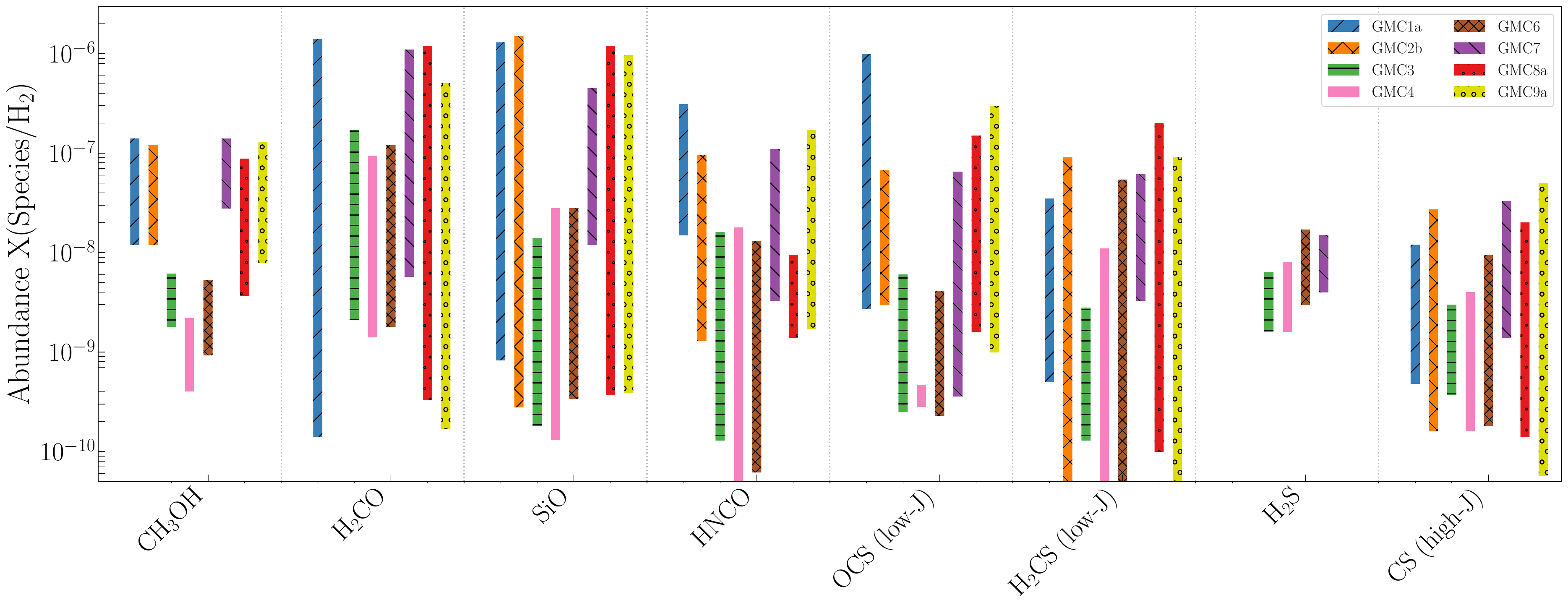}
    \caption{Molecular abundances of CH$_3$OH and H$_2$CO, as well as for the other shock tracers. The values for CH$_3$OH and H$_2$CO are taken from Tables~\ref{tab:Bayesian_gasphys_ch3oh} and \ref{tab:Bayesian_gasphys_h2co}, respectively. The abundances for the other species are displayed in Table~\ref{tab:abundance_range}. }
    \label{fig:abundances}
\end{figure*}

The range of abundances derived for methanol, X$_{\rm{CH3OH}}$, and formaldehyde, X$_{\rm{H2CO}}$, are listed in Tables~\ref{tab:Bayesian_gasphys_ch3oh} and \ref{tab:Bayesian_gasphys_h2co} and shown in Figure~\ref{fig:abundances}. Since the column densities of the E- and A- forms of CH$_3$OH are similar when taking account of the range of uncertainties, we derived X$_{\rm{CH3OH}}$ for both forms together. We did the same for X$_{\rm{H2CO}}$, the p- and o- forms having the same column densities within the range of uncertainties derived. Overall, we derive X$_{\rm{CH3OH}}$ in the range $4.0\times 10^{-10} - 1.4 \times 10^{-7} $, with the lowest abundances towards GMC\,4, and the highest towards the four outermost GMCs, i.e., GMC\,1a, GMC\,2b, GMC\,8a, and GMC\,9a. For X$_{\rm{H2CO}}$, the range is $1.4\times 10^{-10} - 1.4\times10^{-6}$. The range of abundances for H$_2$CO stays relatively constant between the different regions.

\section{Discussion}
\label{sec:Discussion}
In this Section we discuss our results within the framework of determining the physical origin of CH$_3$OH and H$_2$CO and whether they are chemically linked. We first examine whether these two species are tracing shocked gas (Section~\ref{subsec:comp_hnco_sio} and \ref{sec:comp_sulfur}) and then attempt to trace their chemical origin and hence their link (or lack thereof) by examining their abundances (Section~\ref{sec:abundance_met_form}) and by exploiting the well measured E/A ratio of methanol to track its origin.
\begin{figure}
    \centering
    \includegraphics[trim = 5mm 0 12mm 8mmmm,clip=True,width=0.49\textwidth]{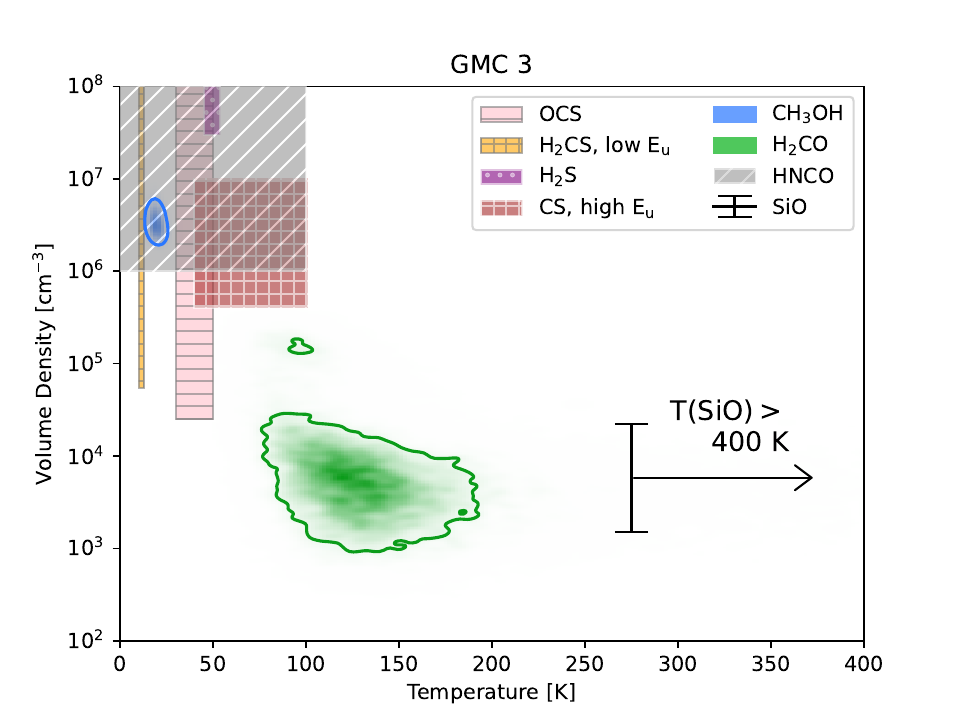}
    \caption{Temperature and volume density ranges toward GMC\,3 for the species studied here, as well as for other species studied using the ALCHEMI survey measurements. Ranges for HNCO and SiO are taken from \cite{Huang+2023}, and ranges for all sulphur-bearing species are taken from \cite{Bouvier+2024}. Note that T(SiO) $>$ 400\,K, so its density range is indicated by the vertical black bar. \ce{CH3OH} and \ce{H2CO} contours encompass the middle 66\% of the temperature and density posterior distributions derived from our Bayesian inference procedure.}
\label{fig:gmc3_temp_dens}
\end{figure}

\subsection{Do methanol and formaldehyde trace the same gas as SiO and HNCO?}\label{subsec:comp_hnco_sio}
The existing evidence for shocks in NGC\,253 includes the detection of HNCO and SiO \citep{GB+2000_sio_253,Meier+2015_hncosio_253, Huang+2023}, the presence of Class I methanol (\ce{CH3OH}) masers \citep{Humire+2022}, and the elevated fractional abundances of \ce{CO2} \citep{Harada+2022}. Given that HNCO and SiO \citep{Huang+2023} were also observed as part of the ALCHEMI large program, we compare the gas characteristics as traced by these two molecules with our \ce{CH3OH} and \ce{H2CO} results. A comparison of these species for GMC\,3 is shown in Figure~\ref{fig:gmc3_temp_dens}, and similar plots for all other GMCs can be found in Appendix \ref{sec:temp_dens_plots}. 
\begin{description}
    \item[CH$_3$OH:] Methanol is tracing denser environments by at least 1-2 orders of magnitude when compared to HNCO (particularly for the outer regions), and by 2-3 orders of magnitude, when compared to SiO for all GMCs.  The temperature traced by \ce{CH3OH}, ranging between $T_{kin} \in 21-36$ K, is the lowest among the tracers, with \ce{HNCO} being slightly higher with a range of $T_{kin} \in 25-100$\,K across all GMCs except for GMC 1a \citep[$T_{kin}\sim230$ K; ][]{Huang+2023}. On the other hand, the presence of methanol masers in the CMZ indicates that the thermal lines of CH$_3$OH could still be associated with shocks. The stark differences in physical properties as traced by SiO and CH$_3$OH lead us to exclude methanol as a tracer of fast shocks. On the other hand, methanol could be tracing slow shocks (which would also explain its abundance in a cold gas). In this scenario, two possibilities could explain the difference in gas properties traced by HNCO and CH$_3$OH. The first is that both species trace the same low-velocity-type of shock, but CH$_3$OH is tracing a denser, and possibly older, gas which would make its cooling timescale faster than that of HNCO. This is consistent with the lower gas temperature traced by CH$_3$OH compared to HNCO. The second possibility is that the two species are tracing shocks of different origins (e.g. sporadic shocks, cloud-cloud collisions or outflow-induced shocks; \citealt{Huang+2023}). Finally, looking at the abundances of \ce{CH3OH}, \ce{SiO}, and \ce{HNCO} towards the CMZ shown in Fig.~\ref{fig:abundances}, we see that while the \ce{CH3OH} abundance clearly decreases towards the innermost GMCs (i.e. GMC\,3, GMC\,4, and GMC\,6), the trend is less clear for the abundances of \ce{SiO} and \ce{HNCO}, although their maximum derived abundance is the lowest towards these same GMCs. This could support the hypothesis of \ce{CH3OH} tracing shocks which may be different to those traced by SiO and HNCO.
    
    \cite{Huang2023FaDi} performed chemical modelling of these three species in NGC 253 and found that a high cosmic ray ionization rate (CRIR) suppresses the abundances of these three species.  It is therefore possible that the general decrease in abundances is partly due to the high cosmic ray ionization rate observed in the inner GMCs \citep{Holdship+2022, Behrens2022ApJ, Behrens2024}. However we also note that  chemical models show that all these species are completely destroyed once the CRIR reaches 1000 time the galactic one.  Hence it is unlikely that the differences we see among CH$_3$OH, SiO and HNCO are primarily due to the enhanced cosmic ray ionization rate. 
    \item[H$_2$CO:] The gas density traced by \ce{H2CO} lies between those derived from \ce{SiO} and \ce{HNCO}. \ce{H2CO} consistently probes gas temperatures around 100\,K except for GMC 4 and GMC 6 where the derived temperature is $>200$ K, although not very well constrained.  H$_2$CO seems therefore to trace a separate gas component that is associated with the heating sources of the GMCs, and not exclusively with shocks. This is also consistent with the fact that contrary to \ce{CH3OH}, the abundance of \ce{H2CO} remains constant throughout the CMZ, as shown in Fig.~\ref{fig:abundances}.
\end{description}

\subsection{Comparison with S-bearing species}
\label{sec:comp_sulfur}

Most S-bearing species have been found to be tracing slow shocks toward the CMZ of NGC 253 (see \citealt{Bouvier+2024}). Comparing the gas parameters traced by methanol with those traced by these S-bearing species (see Figure~\ref{fig:gmc3_temp_dens} and Appendix~\ref{sec:temp_dens_plots}), 
we see in general that the densities and temperatures derived from methanol are close to those derived from the sulfur-bearing species, in particular H$_2$CS, and OCS. This is somewhat consistent with the conclusion that CH$_3$OH may trace slow shocks/post-shock gas. This possibility is also consistent with the fact that Class I methanol masers were detected in the CMZ \citep{Humire+2022}. Interestingly, from Fig.~\ref{fig:abundances}, we can see that the abundance behaviour of \ce{CH3OH} and OCS throughout the GMCs seem to be relatively similar, with a clear decrease towards the innermost GMCs. Looking at the other S-bearing species, their abundance does not show such variation across the GMCs, which could indicate that \ce{CH3OH} and OCS are tracing similar types of shocks.  H$_2$CO, on the other hand, is clearly almost always
"displaced" in density and temperature from the gas traced by sulfur bearing species, again supporting our previous conclusion that it is unlikely to trace the same gas as methanol. 


\subsection{Methanol and formaldehyde abundances}
\label{sec:abundance_met_form}

In this section we briefly discuss the derived abundances for methanol and formaldehyde. In particular we note that the abundance range of methanol across the GMCs is consistent with the large ranges found toward massive young stars: e.g. \cite{vandertak2000} find an abundance of $\sim$ 10$^{-9}$ for the coldest sources and of up to $\sim$ 10$^{-7}$ for the warmest sources. They trace the origin of the methanol to icy mantles and hence propose methanol as an evolutionary indicator.  While we are unable to correlate in a similar manner our abundances to a temperature gradient, we note that there is a distinct gap in abundance between the inner (3, 4 and 6) and outer (1a, 2b, 7, 8a, 9a) GMCs, with the outer regions being richer in methanol. Moreover, abundances in GMCs 3, 4 and 6 are consistent with those found in infrared dark clouds (e.g. Gomez et al. 2011), possibly implying that the inner regions are at a younger evolutionary stage than the outer regions and in agreement with \cite{vandertak2000} conclusions. \cite{vandertak2000} also present H$_2$CO abundances for the same sample of young massive star forming regions and find that H$_2$CO traces cooler gas than methanol, in contrast to our findings.  The derived \cite{vandertak2000} abundances are constant at $10^{-9}$, which is in fact closer to the lower limit for our outer GMCs, with the inner GMCs showing a higher abundance of this species. It is unlikely that we are tracing the same type of gas. Interestingly, they also conclude that H$_2$CO is primarily produced in the gas phase and not correlated to methanol. 

\subsection{The E/A  Ratio in NGC 253 as a probe of methanol formation and destruction}

Methanol is characterized by two spin isomers, $E$ and $A$. The latter has three parallel protons in the methyl group, while in the former one of the protons has an anti-parallel spin compared to the others. Their importance lies in the fact that the conversion of one isomer to the other is highly unlikely within the lifetime of an average methanol molecule (Wirstr{\"o}m et al. 2011).  Hence the ratio of these two isomers should be fixed at the time of the formation of methanol \citep[][]{Kawakita_Kobayashi_2009}. 
In thermal equilibrium (TE)\footnote{This is generally expected for reactions taking place on ice grains. }, the relative population between the E- and A- \ce{CH3OH} is given by:
\begin{equation}
\label{eq:EAR}
    EAR \equiv E/A = \frac{\sum_{i} {g}_{i}^{E-type}\exp(-\frac{{E}_{i}^{E-type}}{k_{B}T_{spin}})}{\sum_{i} {g}_{i}^{A-type}\exp(-\frac{{E}_{i}^{A-type}}{k_{B}T_{spin}})} \in [0.0,1.0], 
\end{equation}
At cold temperatures (e.g. $10-15$ K), EAR $<1.0$ is expected \citep{Friberg1988, Wirstrom+2011}.
If CH$_3$OH is formed at warmer temperatures, the EAR  will reach its theoretical upper limit of 1.0. 
Interestingly, the measured EAR in Galactic star-forming regions is not always consistent with the theoretical predictions \citep[e.g.,][]{Friberg1988, Wirstrom+2011}. Indeed, whilst an EAR $\leq$ 1 has been measured in cold dense cores \citep[e.g.,][]{Friberg1988, Bizzocchi2014, JimenezSerra2021, Megias2023}, Galactic outflows \citep[ e.g.,][]{ Holdship+2019}, and high-mass star-forming regions \citep[e.g.,][]{Menten1988, zhao2023}, EAR > 1  has been measured in some Galactic high-mass star-forming regions \citep[ e.g.,][]{Wirstrom+2011, purcell2009, kalenskii2016, hernandez-hernandez2019} and proto-brown dwarfs \citep{riaz2023}.

From Table \ref{tab:Bayesian_gasphys_ch3oh} we see that, in most GMCs, the overall  isomer ratios, while high, are less than  unity. This suggests that, for these GMCs, methanol has formed on the ices. The kinetic temperatures of the GMCs are generally too high to allow methanol formation on the ices, hence the observed methanol  is a fossil record of a previous colder gas.
However, for the inner GMCs (3, 4 and 6), the EAR is $>$ 1. This may suggest that methanol has formed in the gas phase at higher temperatures. 
Qualitatively this is also consistent with the picture that the CMZ of NGC\,253 is a very dynamic region where large-scale outflows, starburst events, and high cosmic-ray ionization rates \citep[CRIR; e.g. ][]{Holdship+2022,Behrens2022ApJ,Behrens2024} are present. On the other hand, proton exchange reactions with \ce{H3+} with \ce{HCO+} can in principle  equalize the A and E methanol abundances \citep{Wirstrom+2011} if these ions are very abundant, which may be the case in NGC 253 due to the high cosmic ray ionization rate (e.g. \citealt{Holdship+2022}). In this case, methanol can still form in the ice but its EAR would not be fixed at formation.


In extragalactic environments, there are only three other existing E-/A- ratio measurements. 
The EAR inferred by \citet{Muller+2021} for PKS 1830-211 at z=0.89 and the EAR estimated by \citet{Huang+2024} for the nearby galaxy NGC\,1068 are both fairly close to unity. 
Both analyses employed non-LTE radiative transfer modeling. 
On the other hand, the LTE analysis for NGC\,253 by \citet{Humire+2022} gave a wide range of EAR, varying between $\sim1.1-5.0$ \citep[Table 3 in][]{Humire+2022} but with a caveat due to the opacity discussed in their Sect. \ref{sec:radex_phys}. However we note that in the \citet{Muller+2021} and \citet{Huang+2024} studies, the collisional rates for methanol from \citet{Dagdigian_2023} were not employed. 

\section{Conclusions}
\label{sec:conclusions}
We performed a multi-transition study using methanol (\ce{CH3OH}) and formaldehyde (\ce{H2CO}) transitions to constrain the physical properties of the gas at GMC scales ($\sim$ 28\,pc) in the CMZ of NGC\,253 using ALCHEMI data. We conclude the following: 
   \begin{enumerate}
      \item \ce{CH3OH} and \ce{H2CO} trace distinct physical environments. Both molecular species show significant differences in kinetic temperatures and densities. \ce{CH3OH} is found in high-density ($\approx 10^7$cm$^{-3}$) and low-temperature ($T_{kin}<40$~K) regions, while \ce{H2CO} is associated with lower-density ($\approx 10^4$cm$^{-3}$) but warm gas ($T_{kin}>100$~K).  
      The presence of sub-thermal excitation conditions in many \ce{CH3OH} and \ce{H2CO} transitions suggests that different non-thermal desorption mechanisms, such as periodic shocks or cosmic ray-induced desorption, may be responsible for maintaining methanol in the gas phase. This suggests that the formation processes of these two species are not directly linked within NGC\,253's starburst environment.
      


     \item \ce{CH3OH} and \ce{H2CO} do not trace the same gas conditions as classical shock tracers like silicon monoxide (\ce{SiO}) and isocyanic acid (\ce{HNCO}). \ce{CH3OH} is found in significantly denser regions and appears to trace slow shocks or post-shock gas, rather than fast shocks. In contrast, \ce{H2CO} traces a different gas component, likely associated with local heating sources rather than direct shock interactions. 
      
      \item The density and temperature conditions traced by \ce{CH3OH} closely resemble those of sulfur-bearing species like \ce{H2CS} and \ce{OCS}, further supporting its association with slow shocks. The similar abundance trends of \ce{CH3OH} and \ce{OCS} across the GMCs suggest they may trace the same type of shock-driven chemistry, while other sulfur-bearing species show more uniform distributions. \ce{H2CO}, however, remains distinct, reinforcing that it does not trace the same gas as \ce{CH3OH}.
      



      \item Molecular abundances of methanol suggest an evolutionary trend. The \ce{CH3OH} abundance shows a radial gradient, with higher values in the outer GMCs, suggesting different chemical histories or evolutionary stages between the inner and outer CMZ regions. 
      A distinct abundance gap is found between the inner (3, 4, and 6) and outer (1a, 2b, 7, 8a, 9a) GMCs, with the outer regions being methanol-rich. This suggests that the inner GMCs may be at a younger evolutionary stage, similar to infrared dark clouds, and aligns with the idea that \ce{CH3OH} abundance may be an indicator of star-forming region evolution.  

      \item While previous Galactic studies find that \ce{H2CO} traces cooler gas than methanol, our findings indicate the opposite. The derived \ce{H2CO} abundances in NGC 253 suggest that it is primarily formed in the gas phase, rather than through sublimation from icy mantles. 
      This is consistent with the proposed chemical formation pathway via reactions of \ce{CH3} and atomic oxygen. The uniformity of its volume density across the GMCs further supports this interpretation.

      \item The E/A isomer ratio (EAR) of methanol provides insight into its formation mechanisms in NGC 253. In most GMCs, EAR values remain below unity, indicating ice-grain formation at colder temperatures. However, in inner GMCs (3, 4, and 6), EAR $> 1$ suggests gas-phase methanol formation at higher temperatures, possibly influenced by strong outflows, starburst activity, and high cosmic-ray ionization rates. Proton exchange reactions with \ce{H3+} and \ce{HCO+} may further modify the EAR, preventing it from being solely dictated by formation conditions. Compared to other extragalactic environments, the EAR in NGC 253 displays distinct behavior, emphasizing the impact of starburst-driven processes on molecular chemistry.  
      
   \end{enumerate}

\section{Data availability}
\label{sec:data_availability}

All of the integrated intensities for the \ce{CH3OH} and \ce{H2CO} transitions listed in Tables~\ref{tab:table_obsinfo_ch3oh} and \ref{tab:table_obsinfo_h2co} toward the GMC positions listed in Table~\ref{tab:GMC_locations} are available \href{https://doi.org/10.5281/zenodo.15442733}{online}.

\begin{acknowledgements}
      KYH, SV, and MB received funding from the European Research Council (ERC) Advanced Grant MOPPEX 833460.
      KYH acknowledges assistance from Allegro, the European ALMA Regional Center node in the Netherlands. 
      This paper makes use of the following ALMA data: ADS/JAO.ALMA\#2017.1.00161.L, ADS/JAO.ALMA\#2018.1.00162.S. ALMA is a partnership of ESO (represent- ing its member states), NSF (USA) and NINS (Japan), together with NRC (Canada), MOST and ASIAA (Taiwan), and KASI (Republic of Korea), in co-operation with the Republic of Chile. The Joint ALMA Observatory is operated by ESO, AUI/NRAO and NAOJ.
      This research has made use of spectroscopic and collisional data from the EMAA database (\url{https://emaa.osug.fr} and \url{https://dx.doi.org/10.17178/EMAA}). EMAA is supported by the Observatoire des Sciences de l’Univers de Grenoble (OSUG).
\end{acknowledgements}
\bibliographystyle{aa}
\bibliography{aa54156-25}
\newpage

\begin{appendix}
\onecolumn
\section{\ce{CH3OH} and \ce{H2CO} spectra}
\label{sec:spectra}

\begin{figure*}[h!]
  \centering
  \begin{tabular}[b]{@{}p{0.40\textwidth}@{}}
    \centering\includegraphics[width=0.95\linewidth,trim=5mm 18mm 15mm 30mm, clip=True]{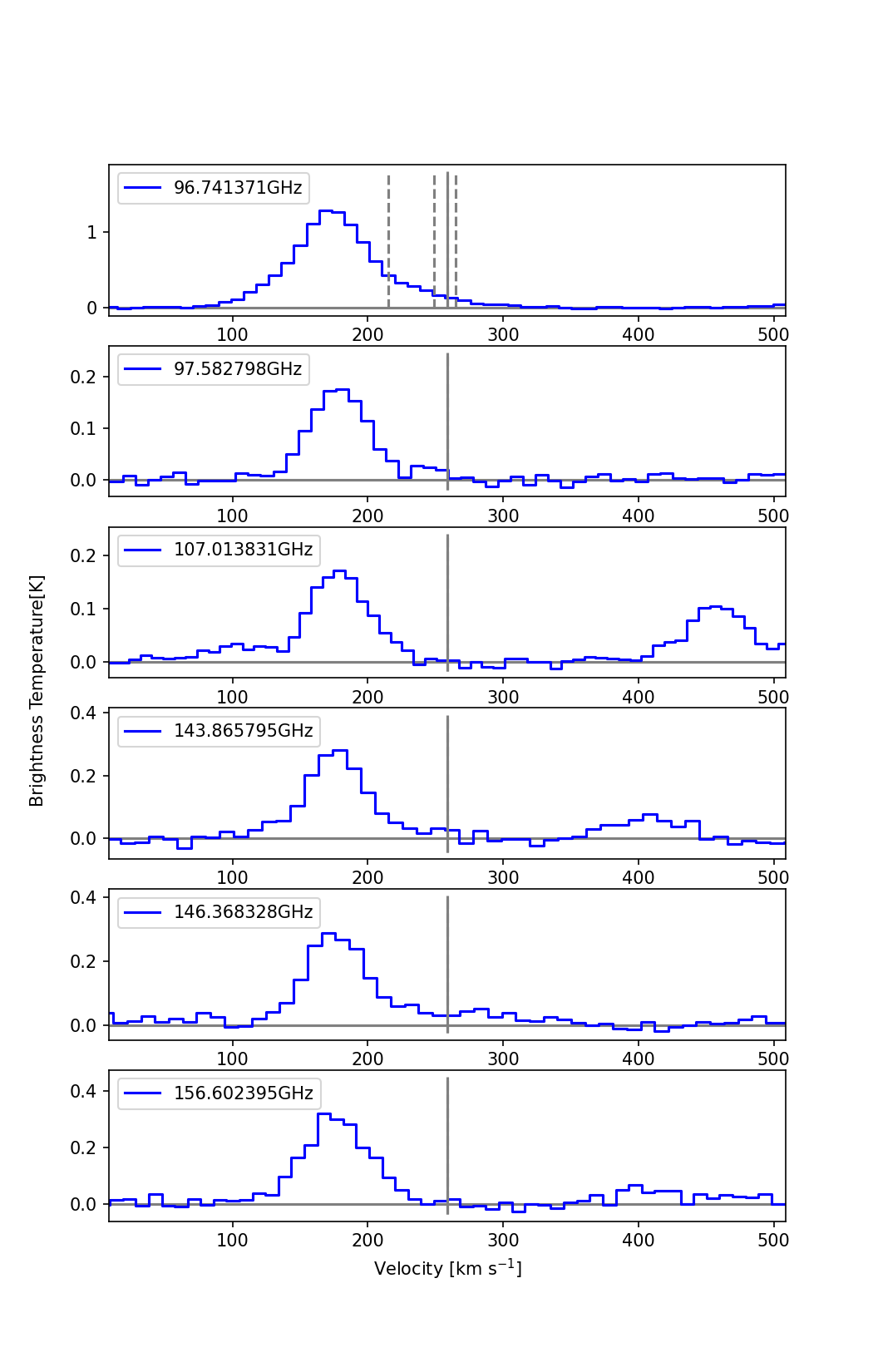} 
  \end{tabular}%
  \quad
  \begin{tabular}[b]{@{}p{0.40\textwidth}@{}}
    \centering\includegraphics[width=0.95\linewidth,trim=5mm 18mm 15mm 30mm, clip=True]{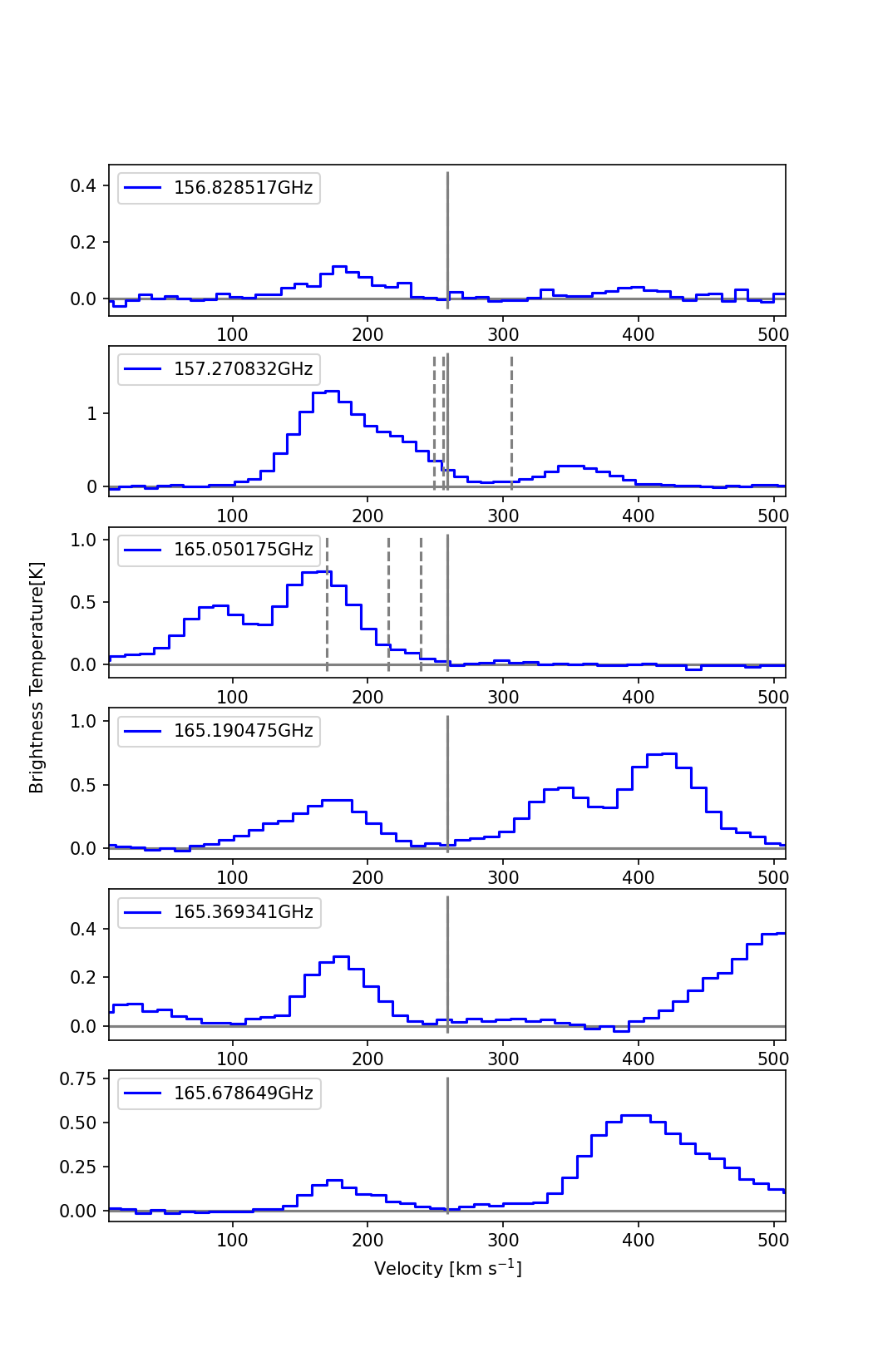} \\
  \end{tabular}
  \begin{tabular}[b]{@{}p{0.40\textwidth}@{}}
    \centering\includegraphics[width=0.95\linewidth,trim=5mm 18mm 15mm 30mm, clip=True]{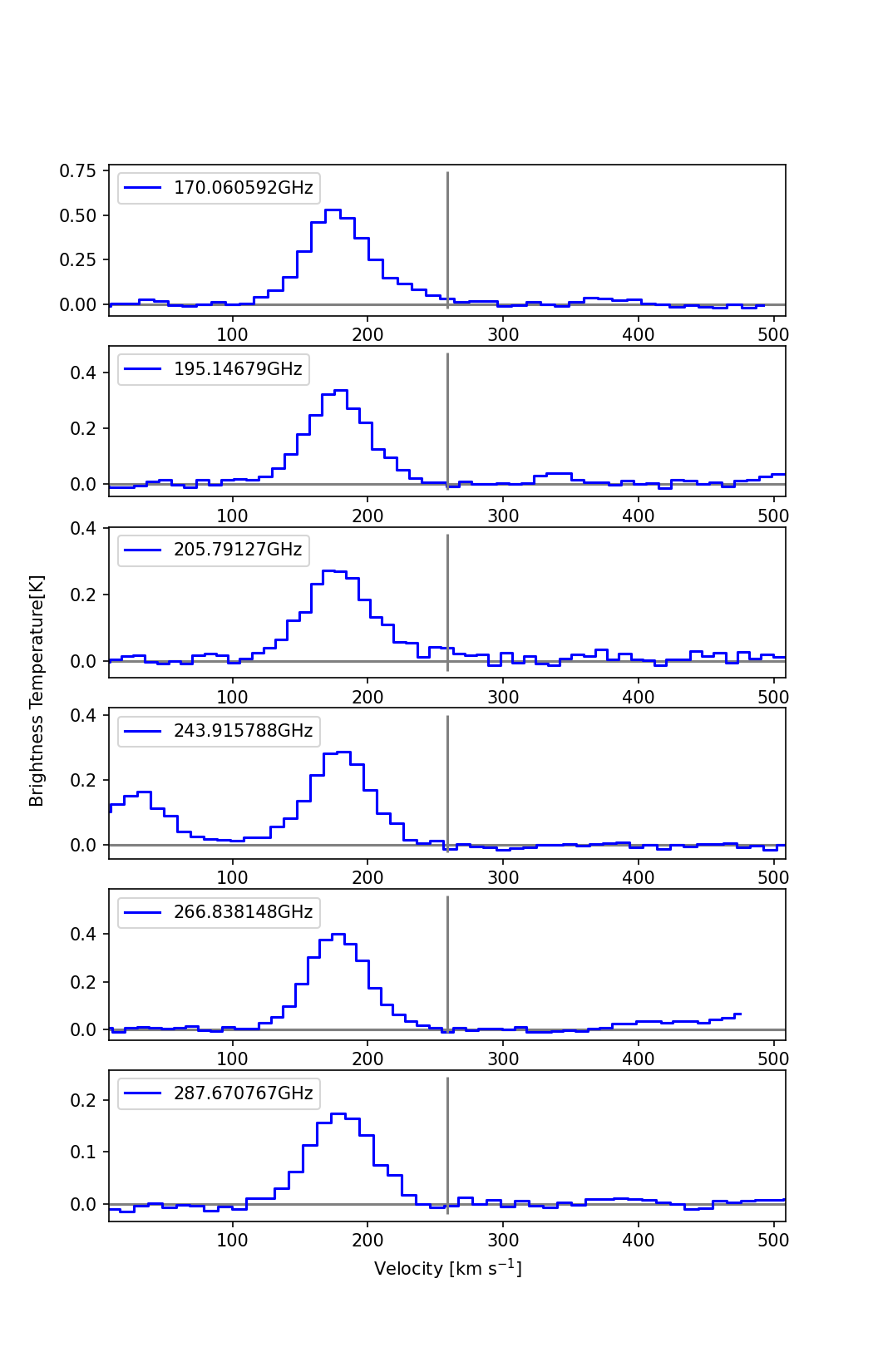} 
  \end{tabular}
  \begin{tabular}[b]{@{}p{0.40\textwidth}@{}}
    \centering\includegraphics[width=0.95\linewidth,trim=5mm 18mm 15mm 30mm, clip=True]{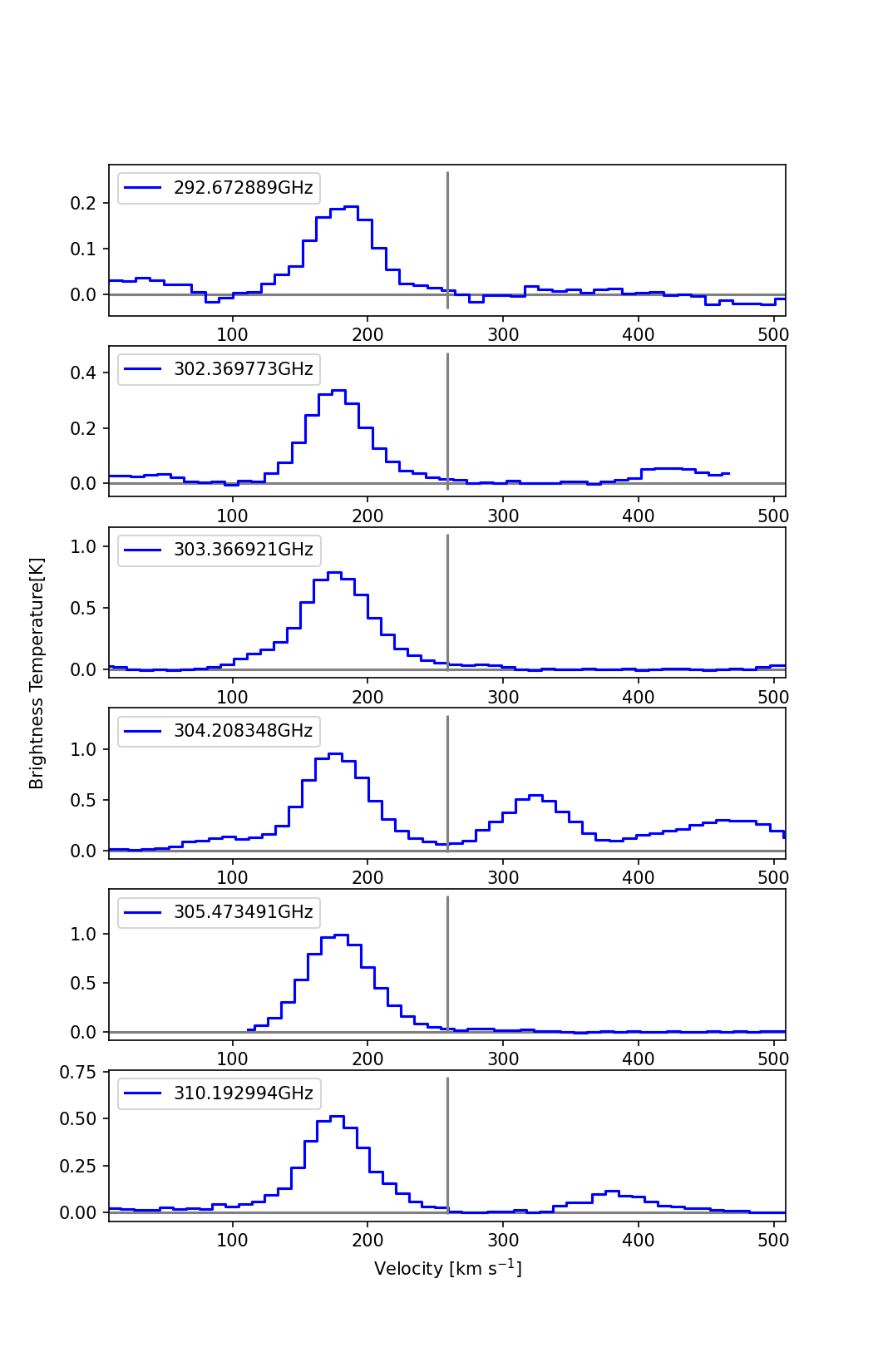} \\
  \end{tabular}
  \caption{The GMC 6 spectra of \ce{CH3OH} transitions investigated in this work. The solid gray lines represent the systemic velocity of NGC\,253 (258\,km\,s$^{-1}$), and the dashed gray lines denote the velocity offsets from Table~\ref{tab:offsets} for each of the blending groups.}
  \label{fig:Spec_I}
\end{figure*}

\begin{figure*}[h!]
  \centering
  \begin{tabular}[b]{@{}p{0.45\textwidth}@{}}
    \centering\includegraphics[width=0.9\linewidth,trim=5mm 18mm 15mm 30mm, clip=True]{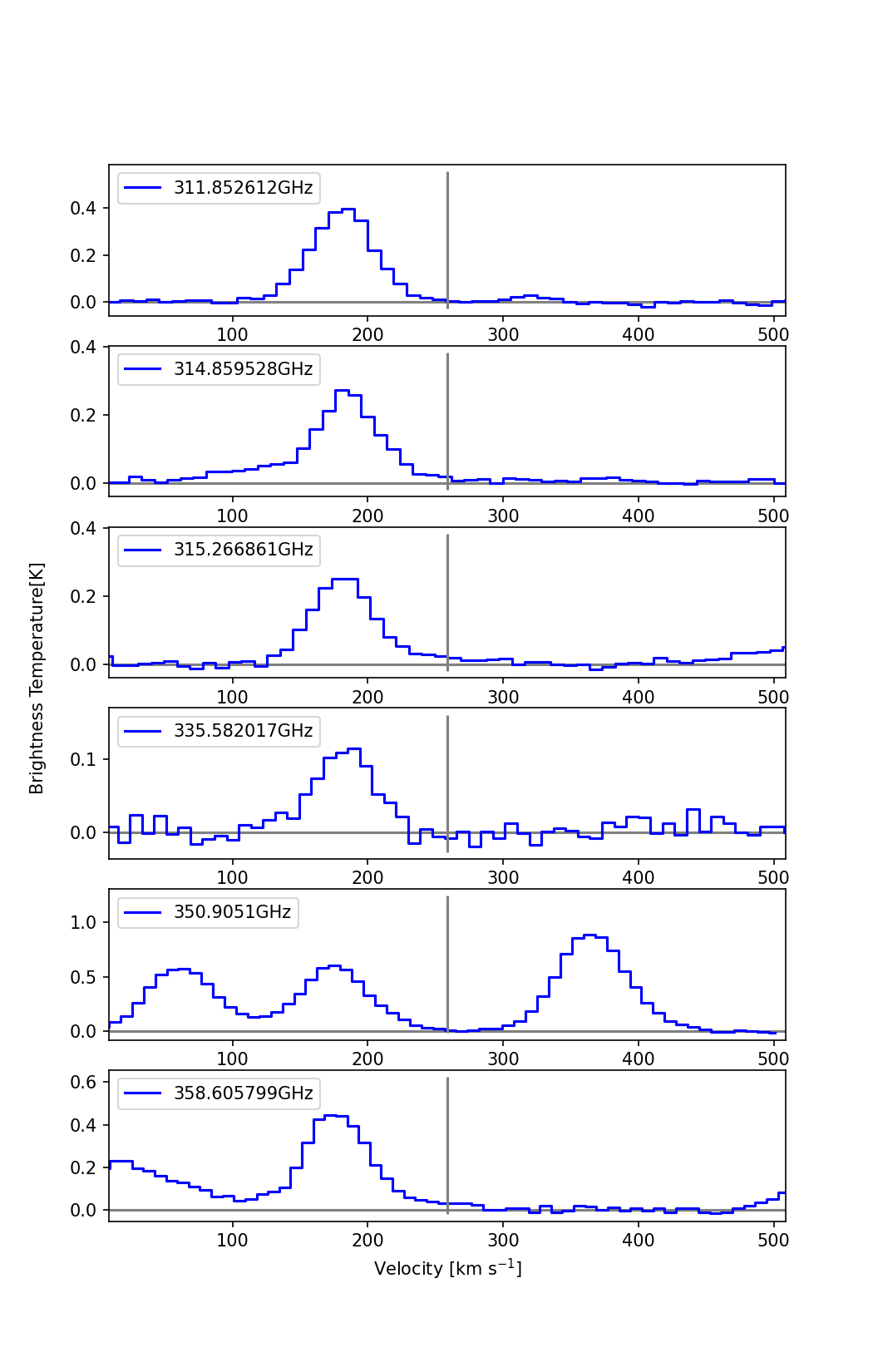} 
  \end{tabular}%
  \quad
  \begin{tabular}[b]{@{}p{0.45\textwidth}@{}}
    \centering\includegraphics[width=0.9\linewidth,trim=5mm 18mm 15mm 30mm, clip=True]{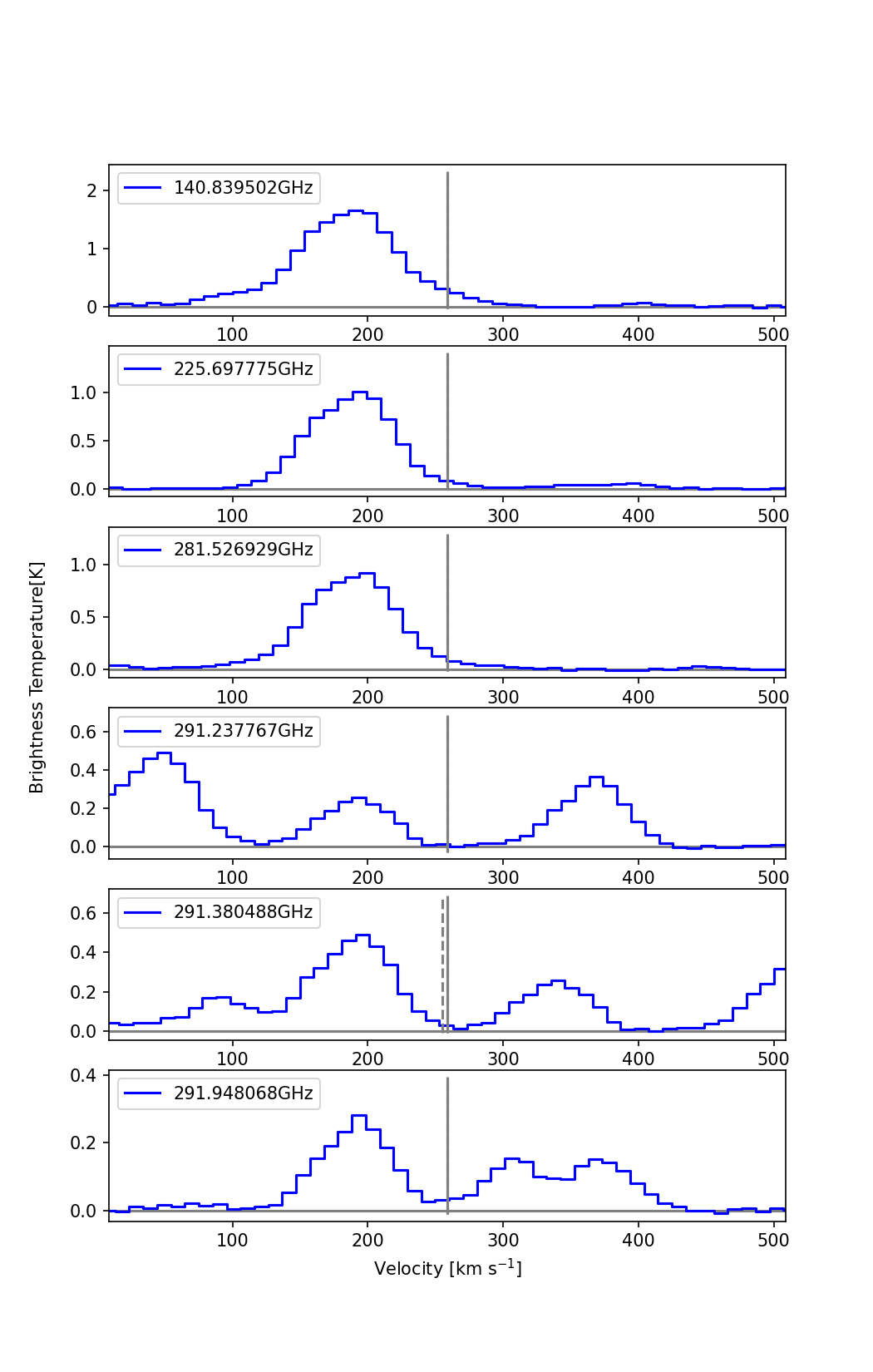} \\
  \end{tabular}
  \begin{tabular}[b]{@{}p{0.45\textwidth}@{}}
    \centering\includegraphics[width=0.9\linewidth,trim=5mm 90mm 15mm 30mm, clip=True]{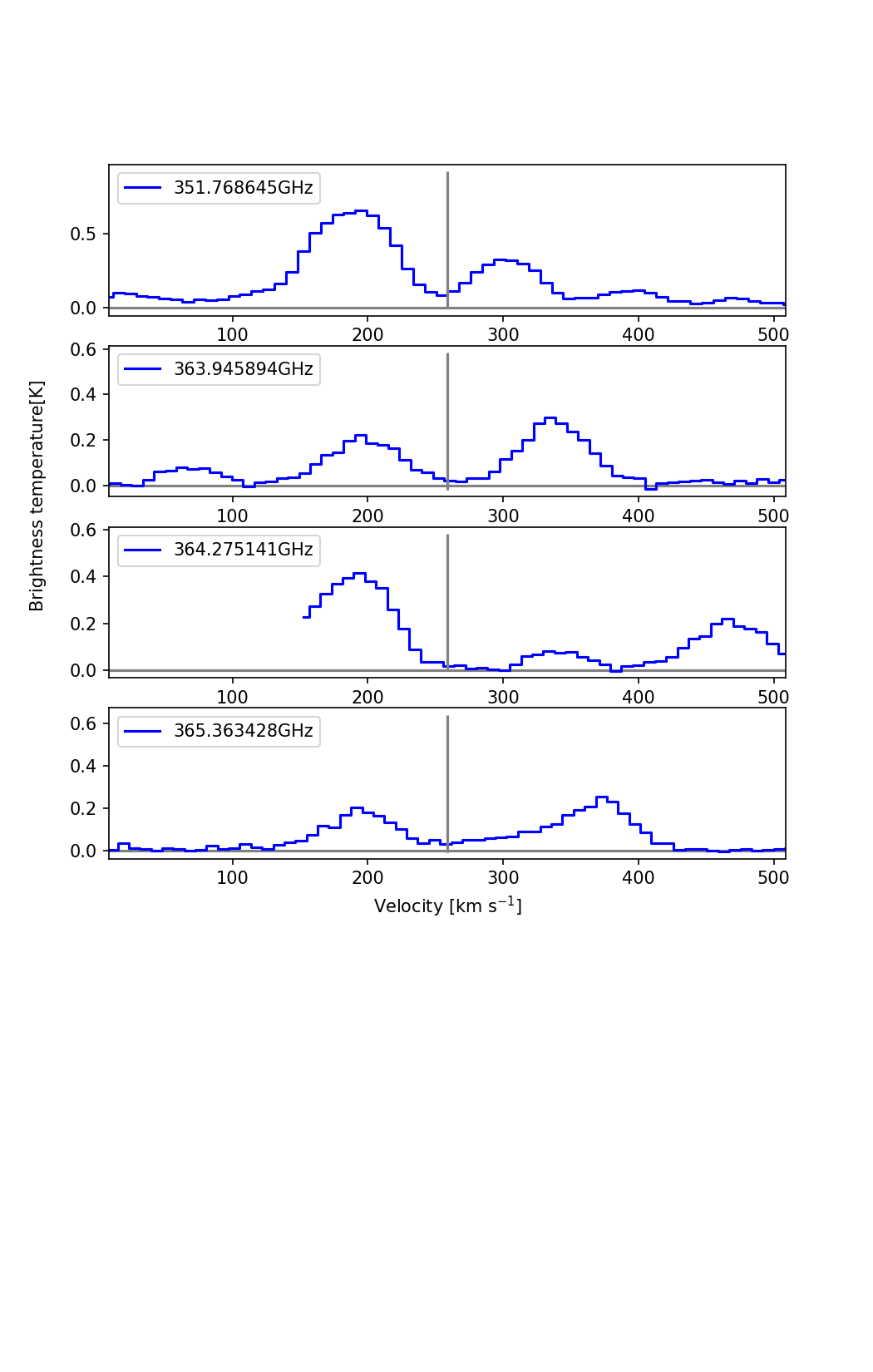} 
  \end{tabular}
  \caption{The GMC 6 spectra of \ce{CH3OH} (upper-left panel; continued from Figure~\ref{fig:Spec_I}) and \ce{H2CO} (upper-right and lower panel) transitions investigated in this work.}
  \label{fig:Spec_II}
\end{figure*}

The \ce{CH3OH} and \ce{H2CO} spectra sampled from GMC 6 are shown in Figures~\ref{fig:Spec_I} and \ref{fig:Spec_II}. Table~\ref{tab:offsets} lists velocity offsets for each \ce{CH3OH} and \ce{H2CO} transition group.

\begin{table*}[h!]
    \centering
    \caption{Velocity offsets for each \ce{CH3OH} and \ce{H2CO} transition group (Figures~\ref{fig:Spec_I} and \ref{fig:Spec_II})}.
    \begin{tabular}{c|ccccc|c}
\hline
Species and Transition Group & Transition & dv \\
&& (km/s) \\
\hline
CH$_3$OH near 96.74\,GHz  & E-\ce{CH3OH} ($2_{-1} - 1_{-1}$) & 6.24 \\
& E-\ce{CH3OH} ($2_0 - 1_0$) & $-9.84$ \\
& E-\ce{CH3OH} ($2_1 - 1_1$) & $-43.79$ \\
& A-\ce{CH3OH} ($2_0 - 1_0$) & 0.0 \\
CH$_3$OH near 157.26\,GHz & E-\ce{CH3OH} ($4_0 - 4_{-1}$) & 47.22 \\
& E-\ce{CH3OH} ($1_0 - 1_{-1}$) & 0.00 \\
& E-\ce{CH3OH} ($3_0 - 3_{-1}$) & $-2.87$ \\
& E-\ce{CH3OH} ($2_0 - 2_{-1}$) & $-9.89$ \\
CH$_3$OH near 165.07\,GHz & E-\ce{CH3OH} ($1_1 - 1_0$) & 0.00 \\
& E-\ce{CH3OH} ($2_1 - 2_0$) & $-19.90$ \\
& E-\ce{CH3OH} ($14_6 - 15_5$) & $-43.92$ \\
& E-\ce{CH3OH} ($3_1 - 3_0$) & $-89.12$ \\
H$_2$CO near 291.38\,GHz & o-\ce{H2CO} ($4_{32} - 3_{31}$) & 0.00 \\
& o-\ce{H2CO} ($4_{31} - 3_{30}$) & $-3.89$ \\
    \hline
    \end{tabular}\\
\label{tab:offsets}
\end{table*}

\FloatBarrier

\section{Moment-0 maps} \label{sec:mom0_maps}
\begin{figure*}[h!]
\centering 
    \includegraphics[scale=0.95]{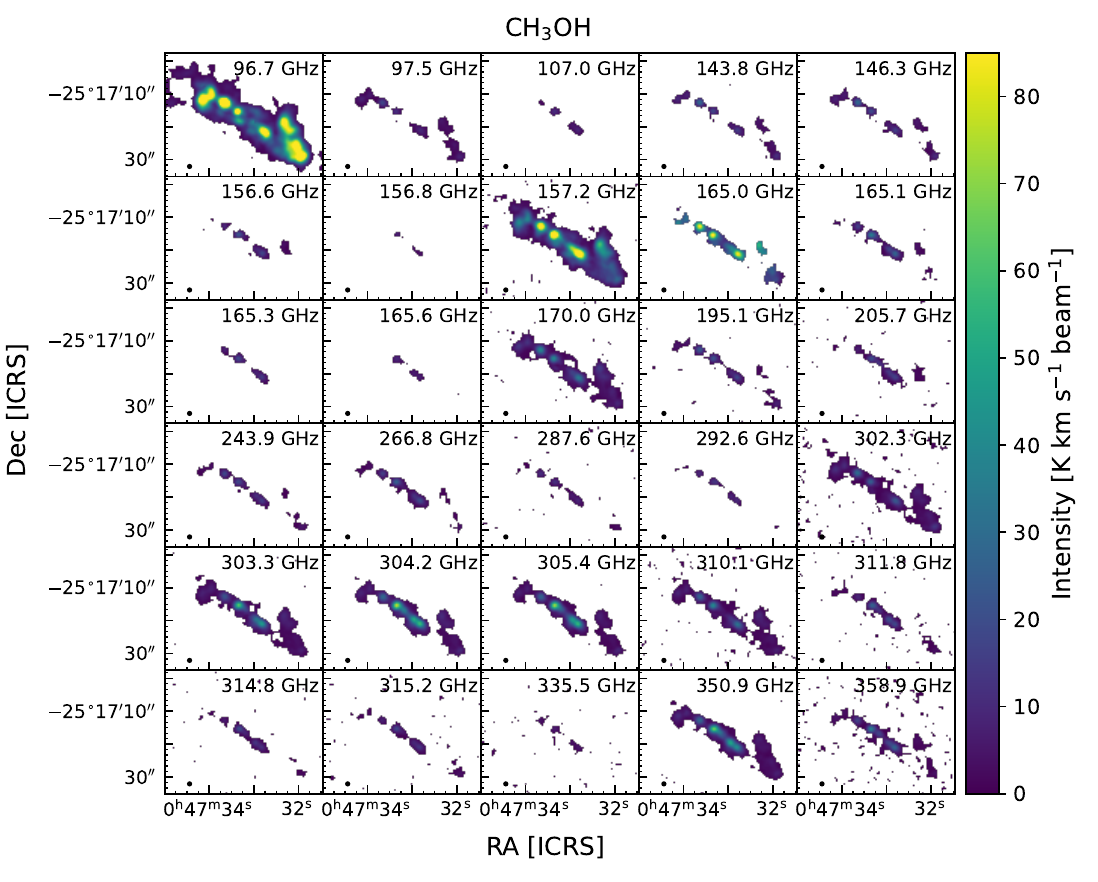}
    \caption{Moment-0 maps for all \ce{CH3OH} transitions listed in Table~\ref{tab:table_obsinfo_ch3oh} with 3$\sigma$ clipping. Black circle in bottom left corner represents the ALCHEMI 1.\!\!$^{\prime\prime}6$/28\,pc beam.}
    \label{fig:all_ch3oh_mom0}
\end{figure*}

\begin{figure*}
    \centering\includegraphics[width=\linewidth]{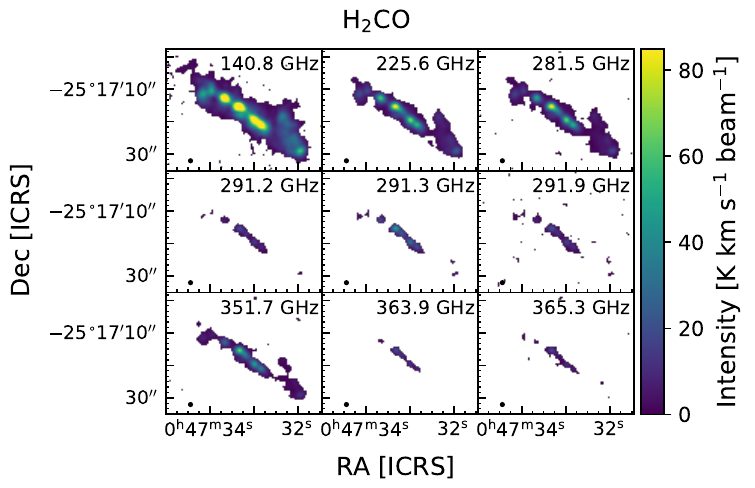}
    \caption{Same as in Figure~\ref{fig:all_ch3oh_mom0} but for \ce{H2CO} with transitions listed in Table~\ref{tab:table_obsinfo_h2co}.}
    \label{fig:all_h2co_mom0}
\end{figure*}

The \ce{CH3OH} and \ce{H2CO} integrated intensity (Moment-0) images for all transitions in this study are shown in Figures~\ref{fig:all_ch3oh_mom0} and \ref{fig:all_h2co_mom0}, respectively.

\FloatBarrier

\clearpage

\section{All corner plots} \label{sec:corners}

\begin{figure*}[h!]
  \centering
  \begin{tabular}[b]{@{}p{0.48\textwidth}@{}}
    \centering\includegraphics[width=0.95\linewidth,trim=3mm 3mm 3mm 0mm, clip=True]{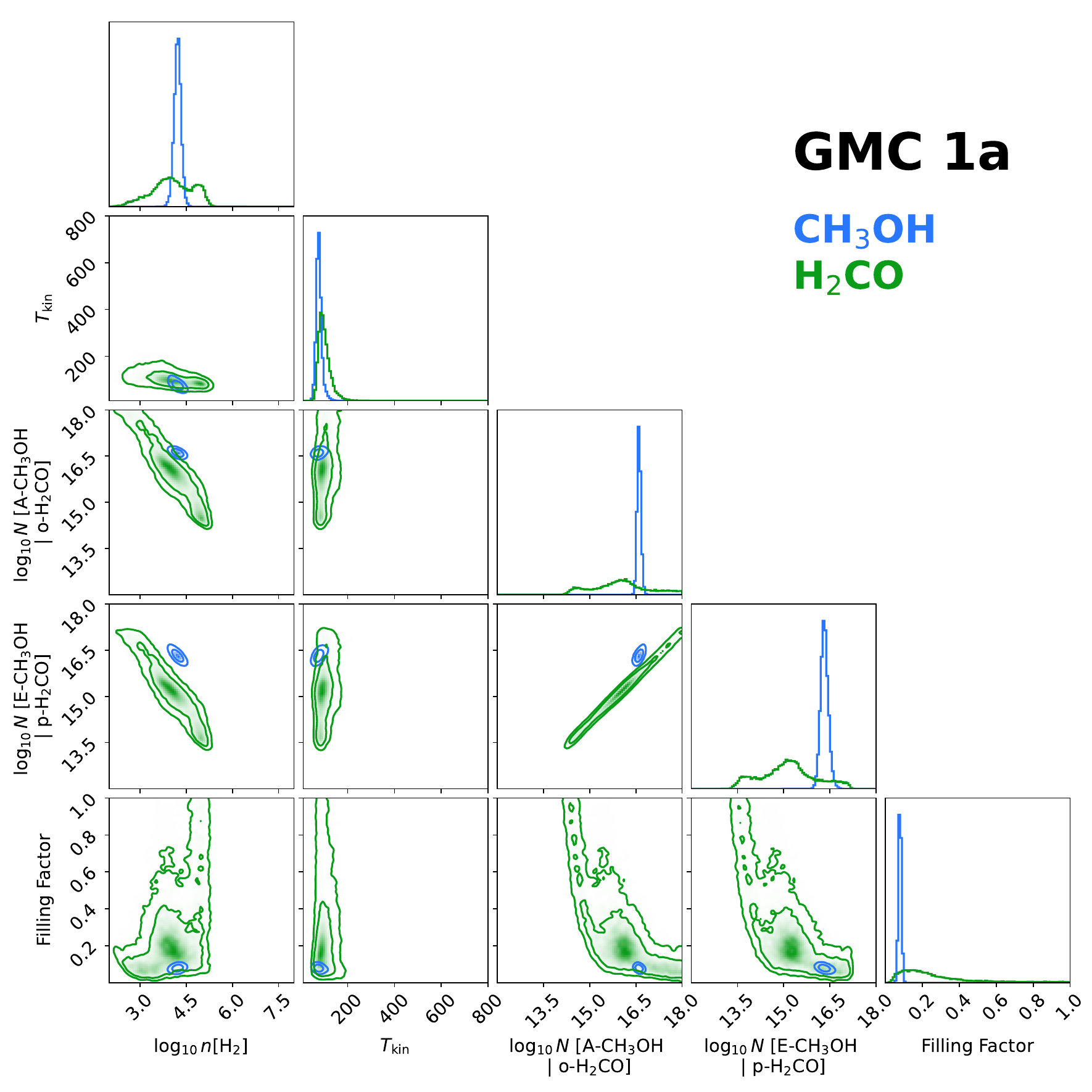} 
  \end{tabular}%
  \quad
  \begin{tabular}[b]{@{}p{0.48\textwidth}@{}}
    \centering\includegraphics[width=0.95\linewidth,trim=3mm 3mm 3mm 0mm, clip=True]{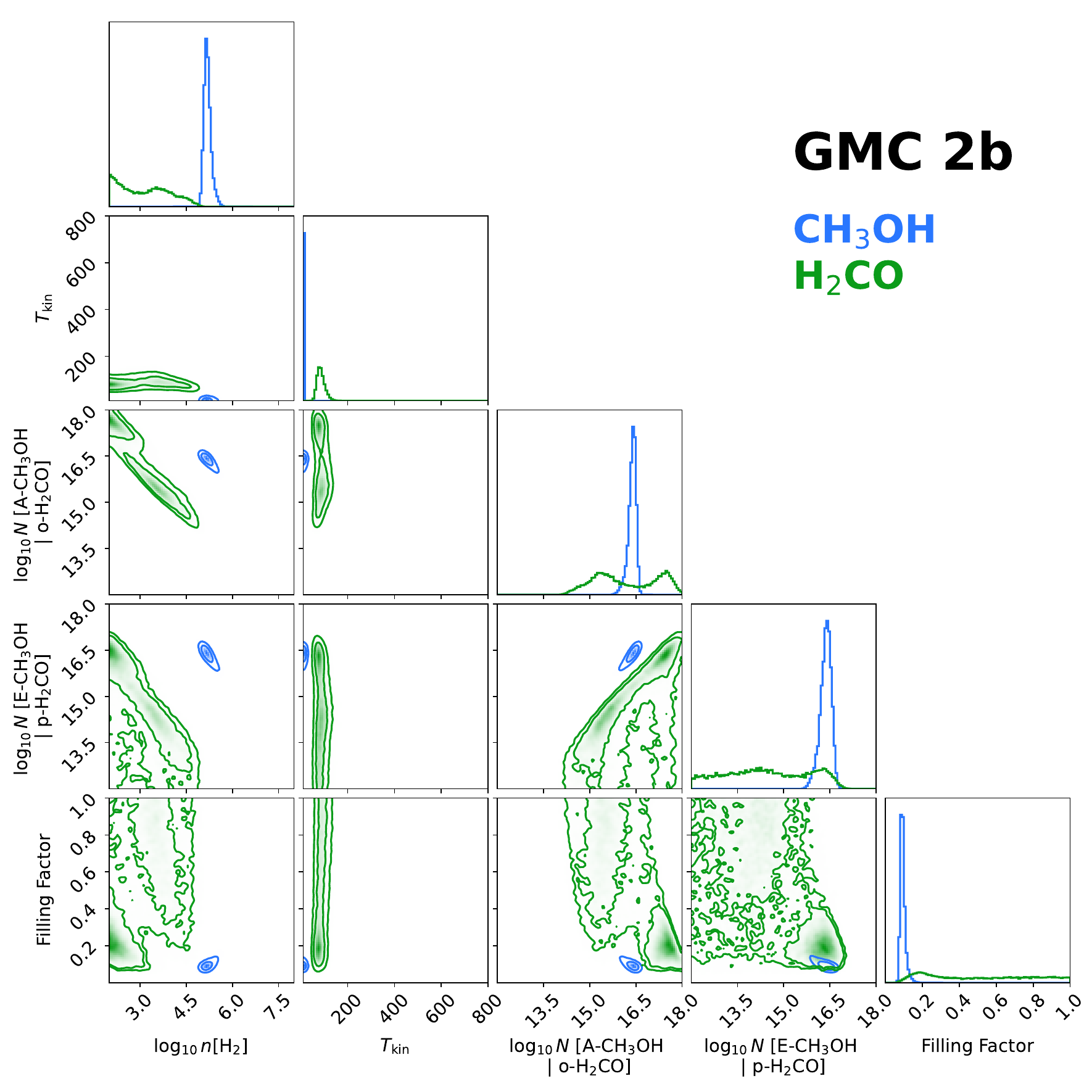} \\
  \end{tabular}
  \begin{tabular}[b]{@{}p{0.48\textwidth}@{}}
    \centering\includegraphics[width=0.95\linewidth,trim=3mm 3mm 3mm 0mm, clip=True]{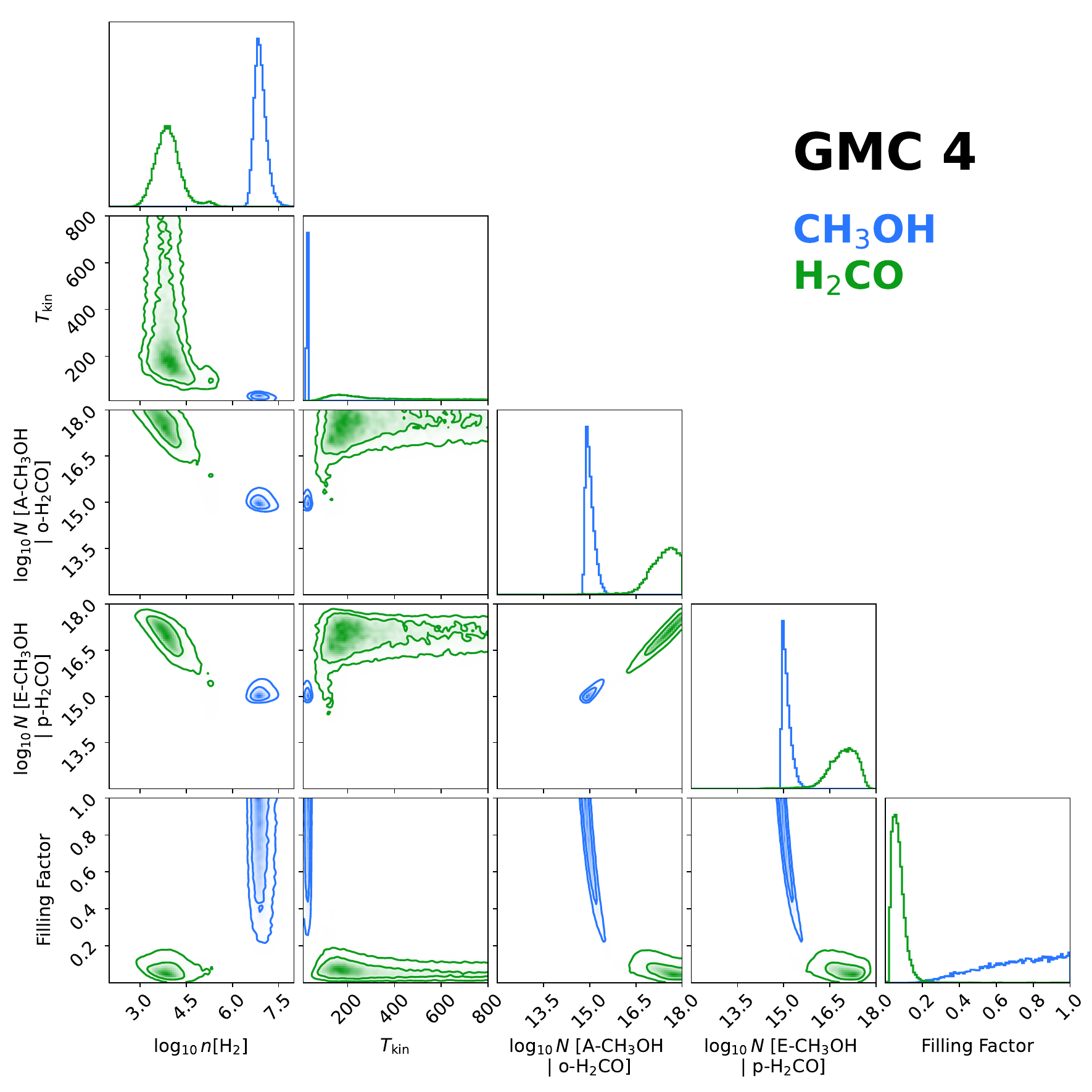} \\
  \end{tabular}
    \begin{tabular}[b]{@{}p{0.48\textwidth}@{}}
    \centering\includegraphics[width=0.95\linewidth,trim=3mm 3mm 3mm 0mm, clip=True]{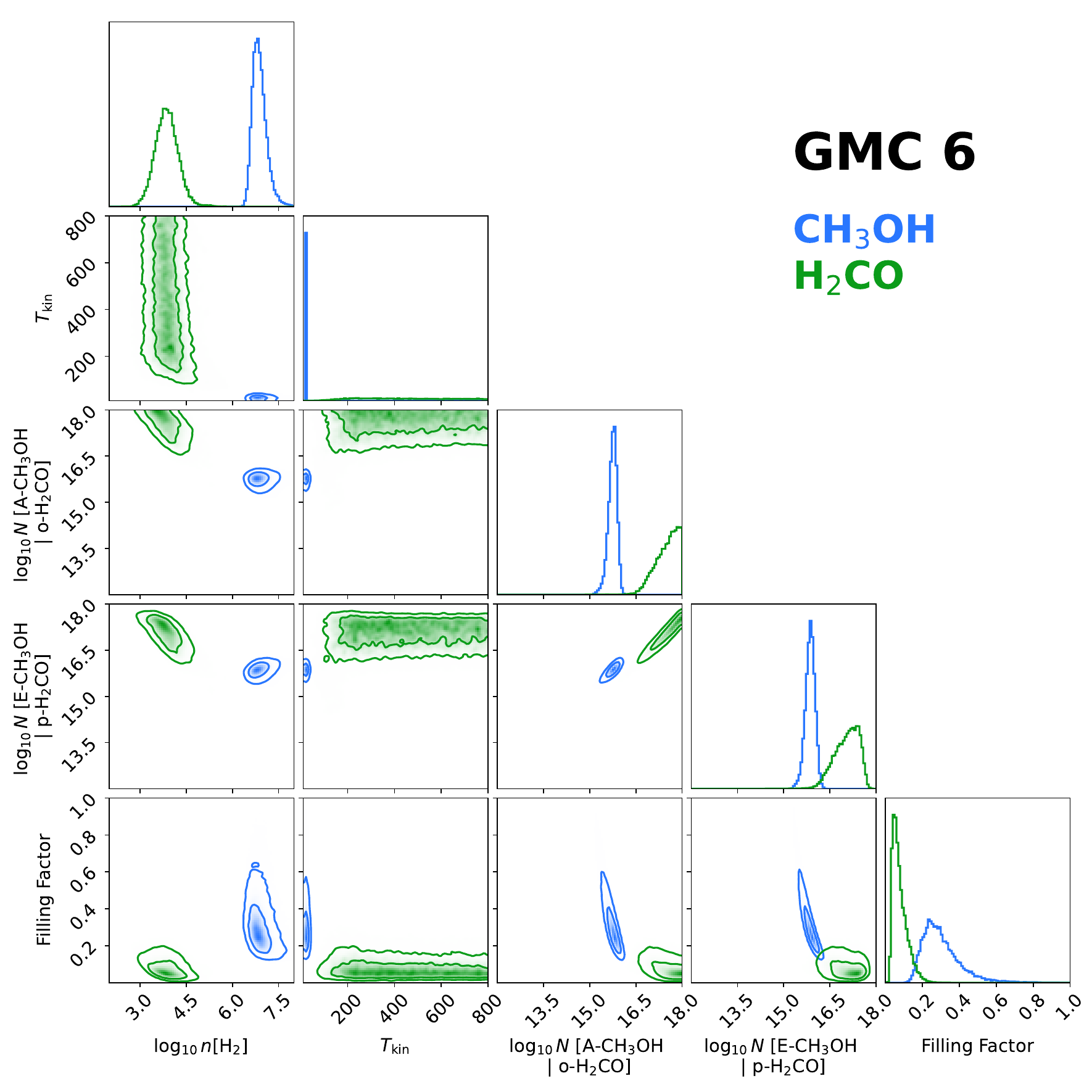} \\
  \end{tabular}
  \caption{Posterior distribution for GMCs 1a, 2b, 4, and 6 obtained using constraints from \ce{CH3OH} (blue) and \ce{H2CO} (green).}
  \label{fig:corners_1a_2b_3_4}
\end{figure*}

\begin{figure*}[h!]
  \centering
  \begin{tabular}[b]{@{}p{0.48\textwidth}@{}}
    \centering\includegraphics[width=0.95\linewidth,trim=3mm 3mm 3mm 0mm, clip=True]{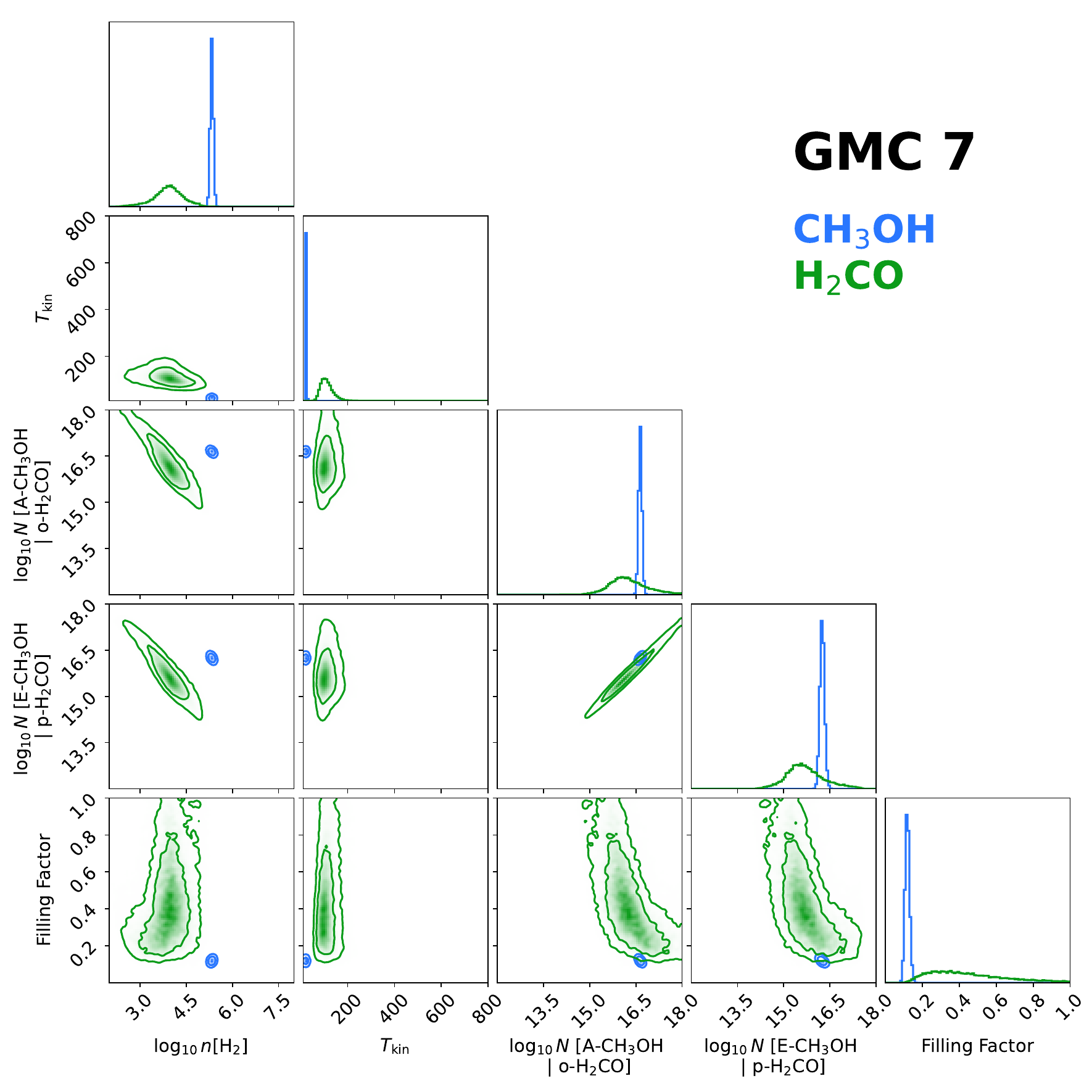} \\
  \end{tabular}
  \begin{tabular}[b]{@{}p{0.48\textwidth}@{}}
    \centering\includegraphics[width=0.95\linewidth,trim=3mm 3mm 3mm 0mm, clip=True]{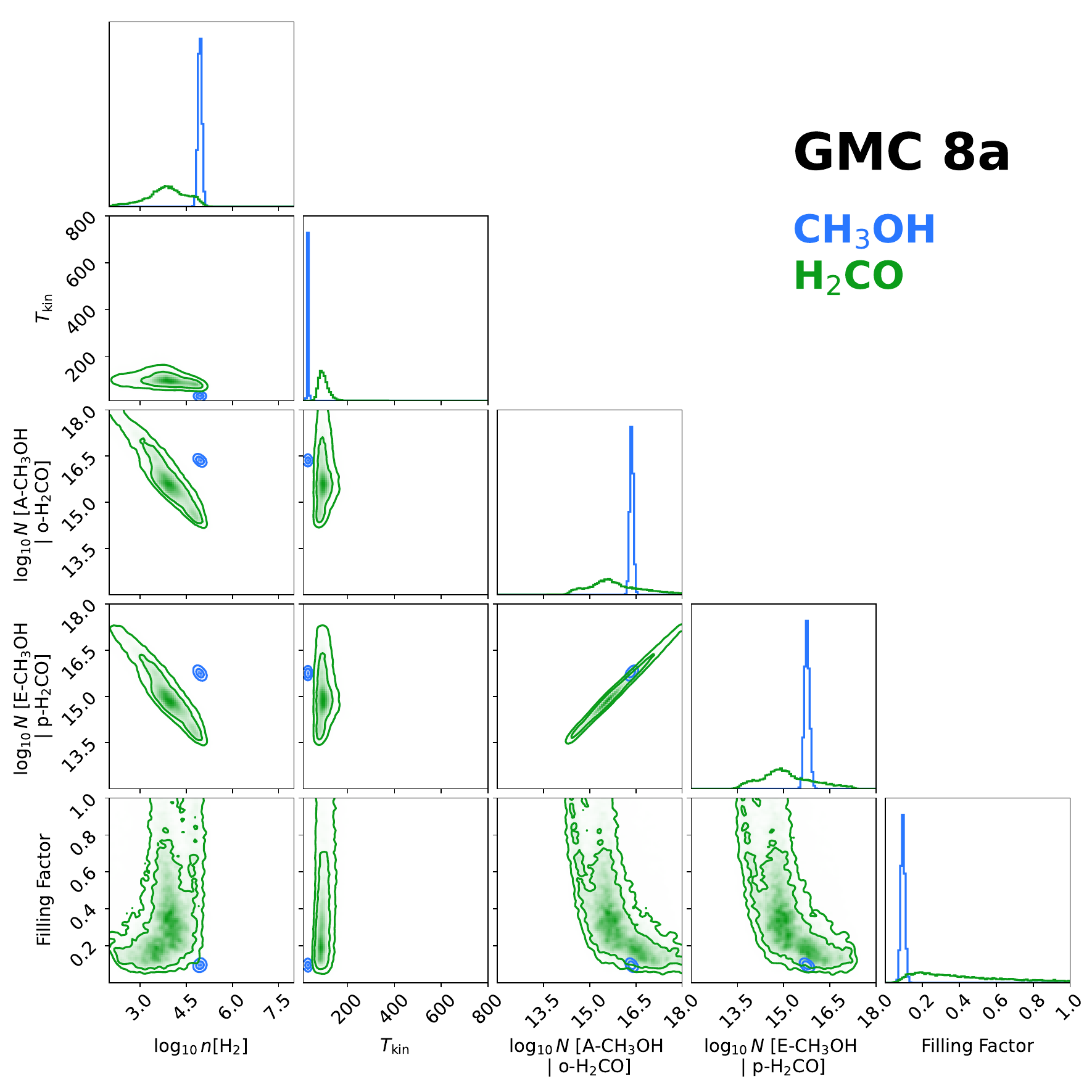} 
  \end{tabular}
  \begin{tabular}[b]{@{}p{0.48\textwidth}@{}}
    \centering\includegraphics[width=0.95\linewidth,trim=3mm 3mm 3mm 0mm, clip=True]{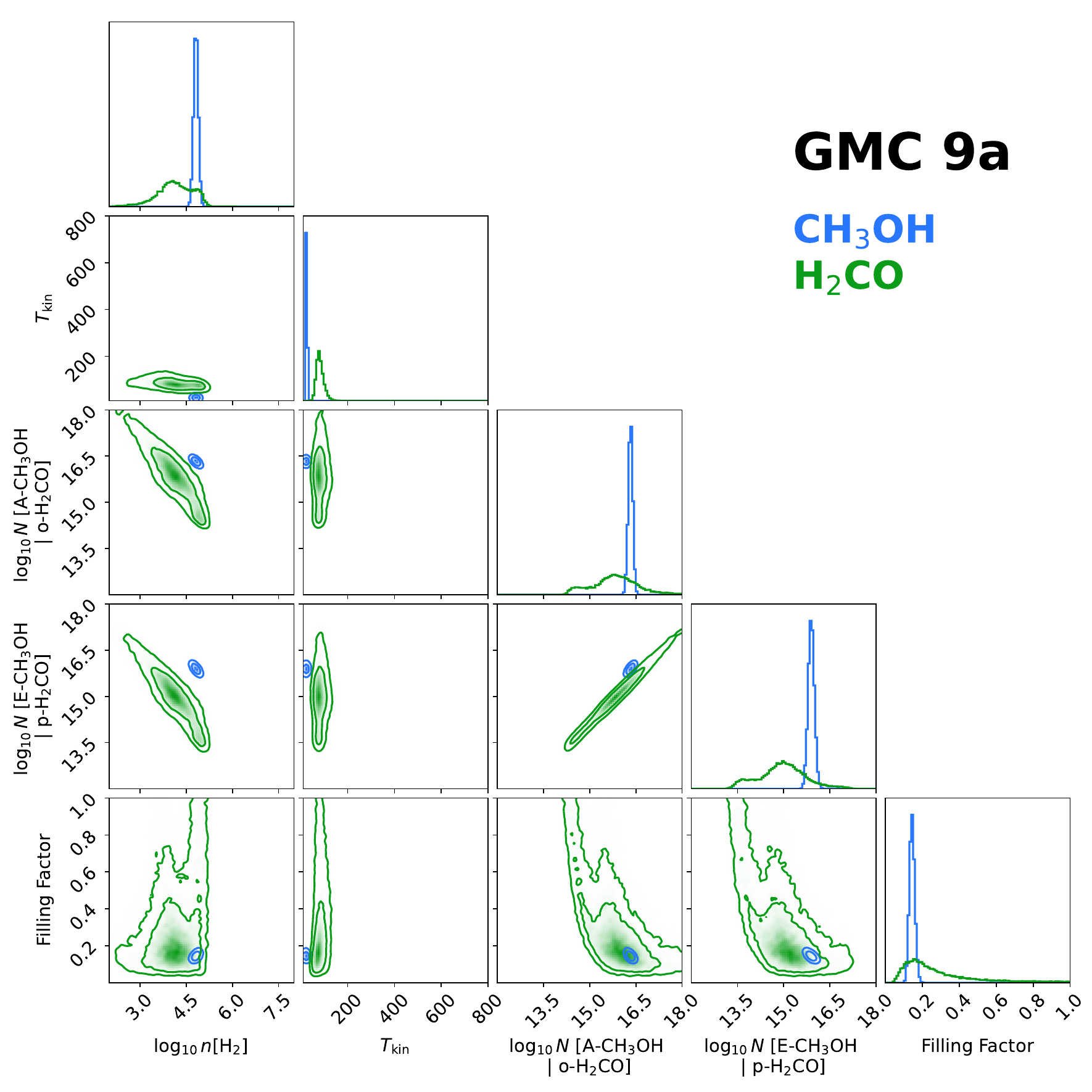} \\
  \end{tabular}
  \caption{Same as in Figure~\ref{fig:corners_1a_2b_3_4} but for GMCs 7, 8a, and 9a.}
  \label{fig:corners_6_7_8a_9a}
\end{figure*}

\ce{CH3OH} and \ce{H2CO} corner plots for GMC\,1a, 2b, 4, and 6 (Figure~\ref{fig:corners_1a_2b_3_4}) and GMCs\,7, 8a, and 9a. (Figure~\ref{fig:corners_6_7_8a_9a}).

\FloatBarrier
\clearpage

\section{Abundances}
We list in Table~\ref{tab:abundance_range} the abundances derived for HNCO, SiO, and the S-bearing species presented in Fig.~\ref{fig:abundances}. To calculate the abundances, we used the column densities of HNCO and SiO from \citet{Huang+2023} and of the S-bearing species from \citet{Bouvier+2024}. We used the same H$_2$ column densities as for CH$_3$OH and H$_2$CO, and we applied the beam filling factor derived for each region and each species from the non-LTE analyses performed in \citet[][see their Table 5]{Huang+2023} and \citet[][see their Tables 3 and C.1]{Bouvier+2024}.  The beam filling factor was calculated from the source's size using the formula for the beam filling factor (Section~\ref{sec:radex}).

\begin{table*}[h!]
    \caption{Abundance ranges derived for HNCO, SiO, and the S-bearing species shown in Fig.~\ref{fig:abundances}.}
    \label{tab:abundance_range}
    \resizebox{\linewidth}{!}{\begin{tabular}{lcccccc}
    \hline
       Region  & X(SiO) & X(HNCO) & X(OCS) & X(H$_2$CS) & X(H$_2$S) & X(CS)  \\
       \hline
        GMC\,1a &$8.3\times 10^{-10}-1.3\times10^{-6}$ & $1.5\times 10^{-8}-3.1\times10^{-7}$& $2.7\times 10^{-9}-1.0\times10^{-7}$& $5.0\times 10^{-10}-3.5\times10^{-8}$& ...& $4.8\times 10^{-10}-1.2\times10^{-8}$\\
        GMC\,2b &$2.8\times 10^{-10}-1.5\times10^{-6}$ & $1.3\times 10^{-9}-9.5\times10^{-8}$& $3.0\times 10^{-9}-6.7\times10^{-8}$& $1.5\times 10^{-11}-9.0\times10^{-8}$& ...& $1.6\times 10^{-10}-2.7\times10^{-10}$ \\
        GMC\,3 &$1.8\times 10^{-10}-1.4\times10^{-8}$ & $1.3\times 10^{-10}-1.6\times10^{-8}$& $2.5\times 10^{-10}-6.0\times10^{-9}$& $1.3\times 10^{-10}-2.8\times10^{-9}$& $1.6\times10^{-9}-6.4\times10^{-9}$& $3.7\times 10^{-10}-3.0\times10^{-9}$ \\
        GMC\,4 &$1.3\times 10^{-10}-2.8\times10^{-8}$ & $1.1\times 10^{-11}-1.8\times10^{-8}$& $2.8\times 10^{-10}-4.7\times10^{-10}$& $2.9\times 10^{-12}-1.1\times10^{-8}$& $1.6\times10^{-9}-8.1\times10^{-9}$& $1.6\times 10^{-10}-4.0\times10^{-9}$ \\
        GMC\,6 &$3.4\times 10^{-10}-2.8\times10^{-8}$ & $6.2\times 10^{-11}-1.3\times10^{-8}$& $2.3\times 10^{-10}-4.1\times10^{-9}$& $4.3\times 10^{-11}-5.4\times10^{-8}$& $3.0\times10^{-9}-1.7\times10^{-8}$& $1.8\times 10^{-10}-9.5\times10^{-9}$ \\
        GMC\,7 &$1.2\times 10^{-8}-4.5\times10^{-7}$ & $3.3\times 10^{-9}-1.1\times10^{-7}$& $3.6\times 10^{-10}-6.5\times10^{-8}$& $3.3\times 10^{-9}-6.2\times10^{-8}$& $4.0\times10^{-9}-1.5\times10^{-8}$& $1.4\times 10^{-9}-3.3\times10^{-8}$ \\
        GMC\,8a &$3.7\times 10^{-10}-1.2\times10^{-6}$ & $1.4\times 10^{-9}-9.5\times10^{-9}$& $1.6\times 10^{-9}-1.5\times10^{-7}$& $1.0\times 10^{-10}-2.0\times10^{-7}$& ...& $1.4\times 10^{-10}-2.0\times10^{-8}$ \\
        GMC\,9a &$3.9\times 10^{-10}-9.7\times10^{-7}$ & $1.7\times 10^{-9}-1.7\times10^{-7}$& $1.0\times 10^{-9}-3.0\times10^{-7}$& $4.0\times 10^{-11}-9.0\times10^{-8}$& ...& $5.7\times 10^{-11}-5.0\times10^{-8}$ \\
         \hline
    \end{tabular}}
\end{table*}

\FloatBarrier

%


\clearpage

\section{Temperature versus density plots} \label{sec:temp_dens_plots}

Kinetic temperature versus volume density plots showing the same information as Figure~\ref{fig:gmc3_temp_dens} but for GMCs\,1a, 2b, 4, 6, 7, 8a, and 9a.
\begin{figure*}[ht!]
  \centering
  \begin{tabular}[b]{@{}p{0.35\textwidth}@{}}
    \centering\includegraphics[width=0.98\linewidth,trim = 5mm 2 15mm 8mm,clip=True]{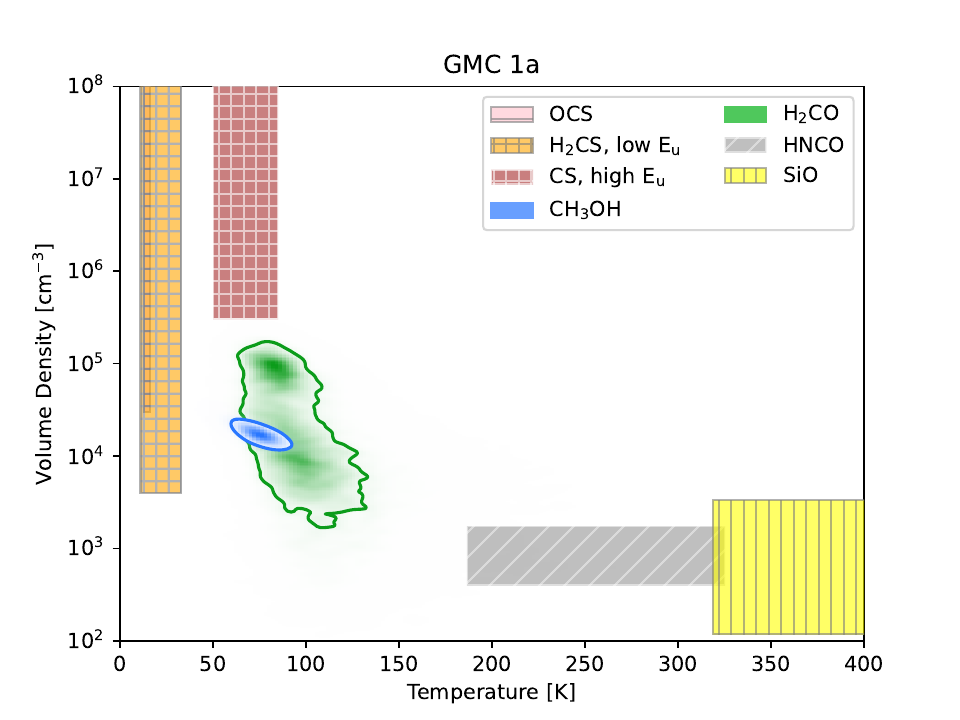} 
  \end{tabular}%
  \quad
  \begin{tabular}[b]{@{}p{0.35\textwidth}@{}}
    \centering\includegraphics[width=0.98\linewidth,trim = 5mm 2mm 15mm 8mm,clip=True]{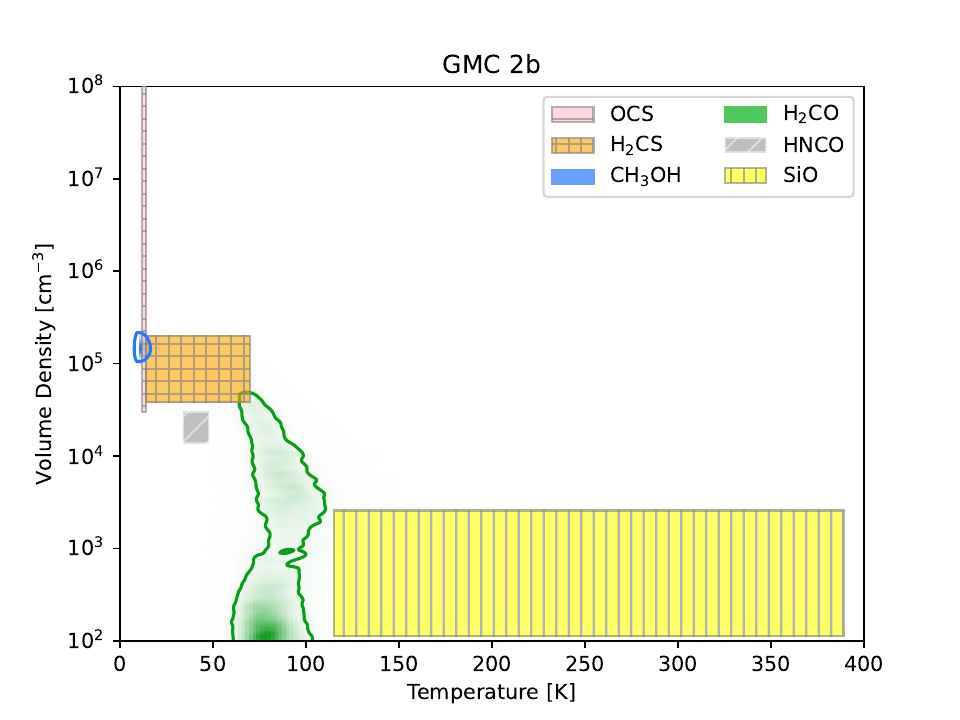} \\
  \end{tabular}
  \begin{tabular}[b]{@{}p{0.35\textwidth}@{}}
    \centering\includegraphics[width=0.98\linewidth,trim = 5mm 2mm 15mm 8mm,clip=True]{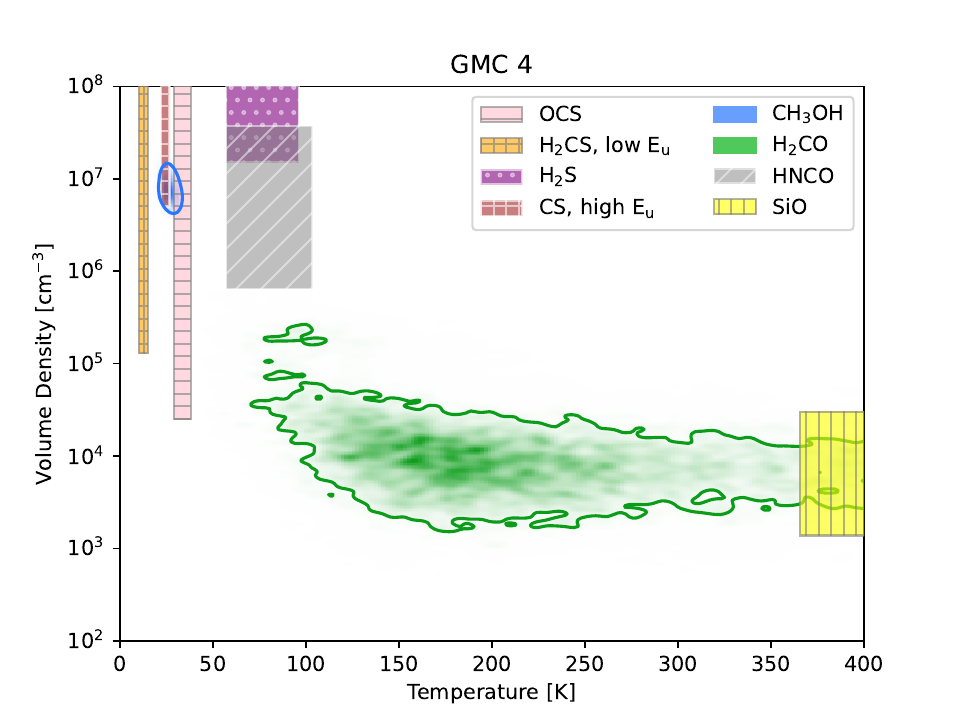} 
  \end{tabular}
  \quad
  \begin{tabular}[b]{@{}p{0.35\textwidth}@{}}
    \centering\includegraphics[width=0.98\linewidth,trim = 5mm 2mm 15mm 8mm,clip=True]{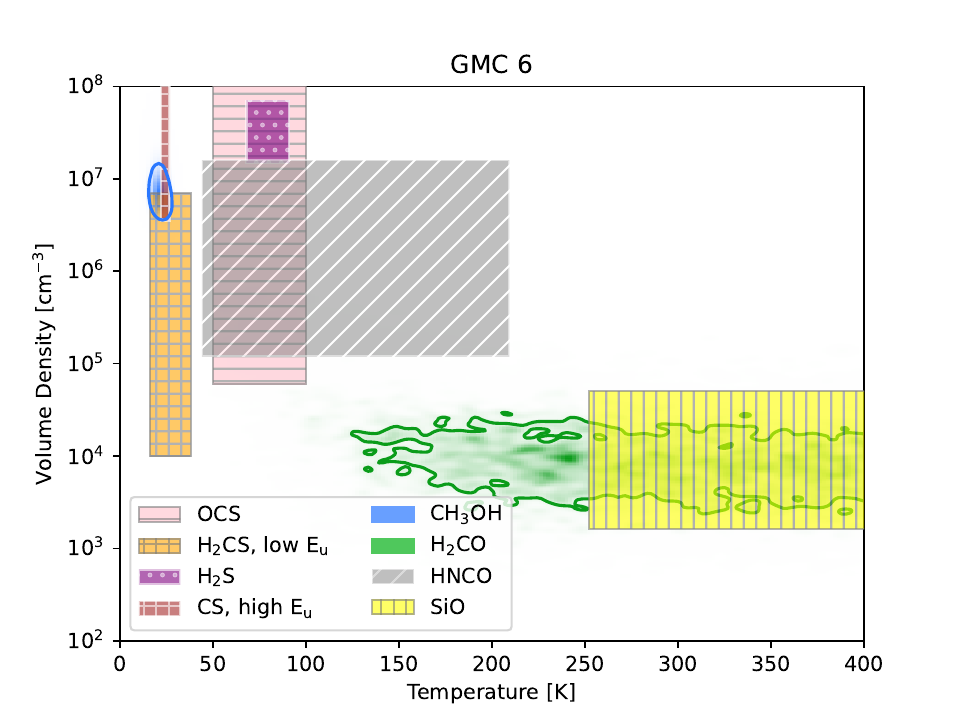} \\
  \end{tabular}

  \begin{tabular}[b]{@{}p{0.35\textwidth}@{}}
    \centering\includegraphics[width=0.98\linewidth,trim = 5mm 2mm 15mm 8mm,clip=True]{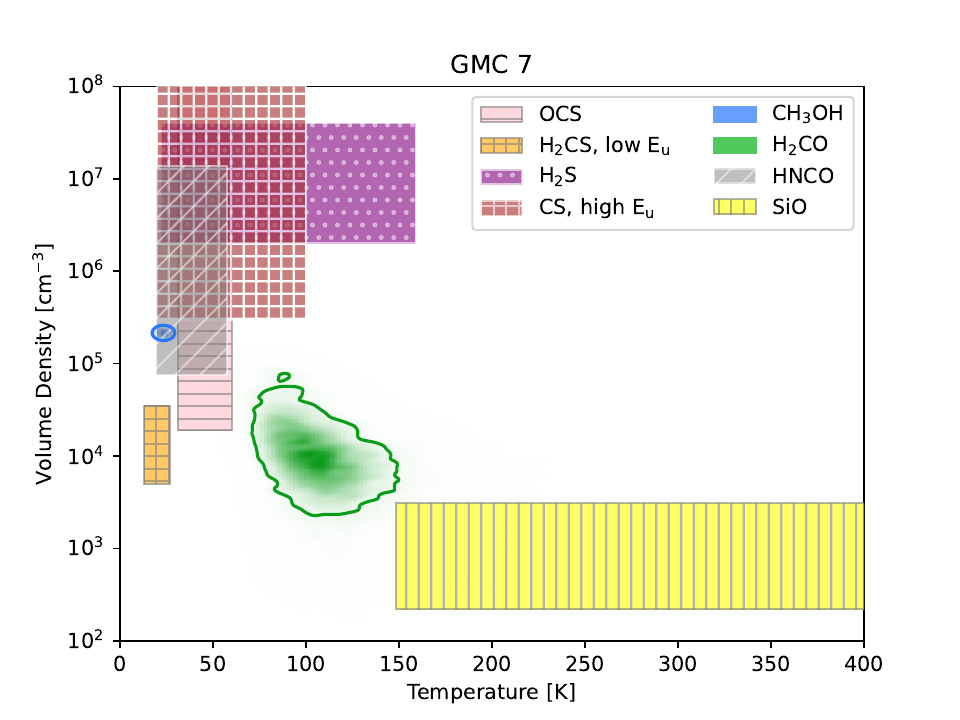} 
  \end{tabular}%
  \quad
  \begin{tabular}[b]{@{}p{0.35\textwidth}@{}}
    \centering\includegraphics[width=0.98\linewidth,trim = 5mm 2mm 15mm 8mm,clip=True]{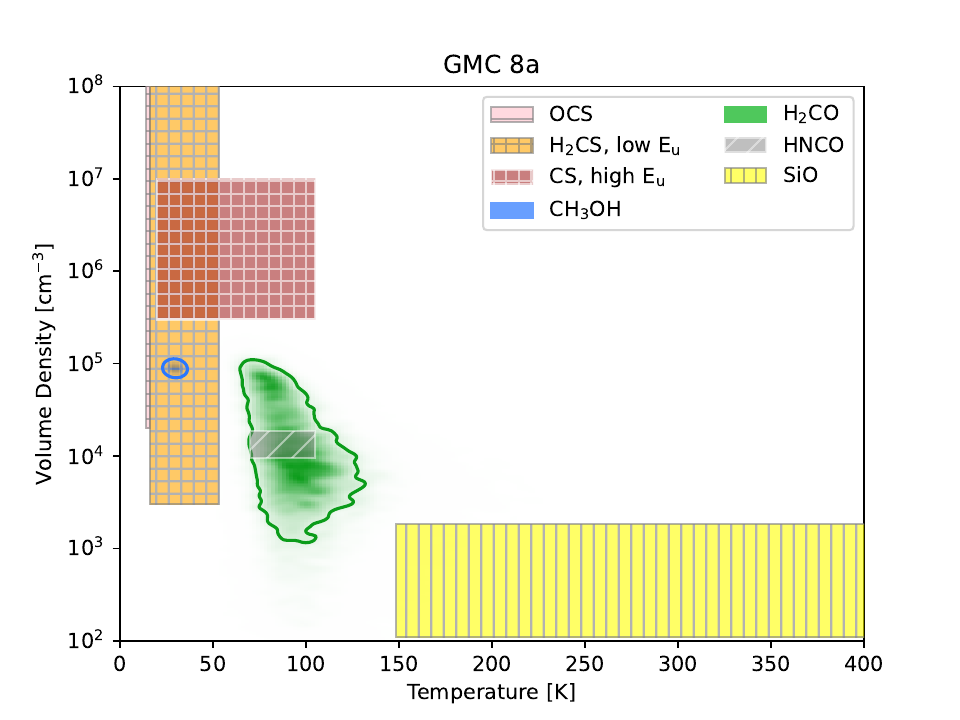} \\
  \end{tabular}
  \begin{tabular}[b]{@{}p{0.35\textwidth}@{}}
    \centering\includegraphics[width=0.98\linewidth,trim = 5mm 2mm 15mm 8mm,clip=True]{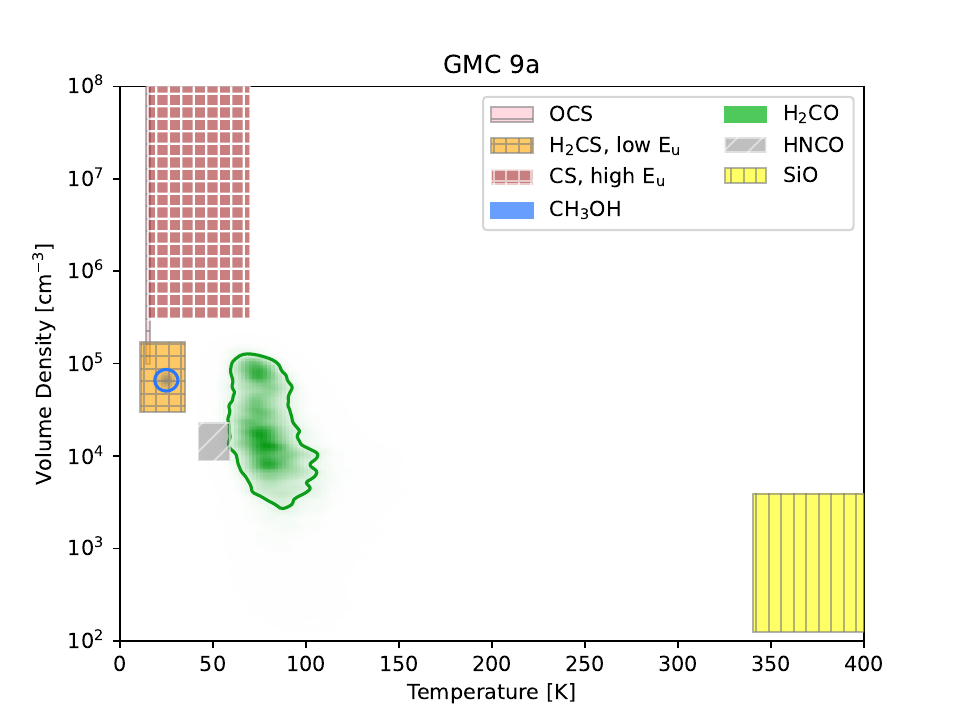} 
  \end{tabular}
    \caption{Same as in Figure~\ref{fig:gmc3_temp_dens} but for GMCs 1a, 2b, 4, 6, 7, 8a, and 9a.}
  \label{fig:temp_dens_appendix}
\end{figure*}

\end{appendix}
%
%
\label{LastPage}
\end{document}